\documentclass[reqno]{amsart}
\usepackage{times}
\usepackage{epsfig}
\usepackage{graphicx}
\usepackage{amsmath}
\usepackage{amssymb}
\usepackage{subfigure}
\usepackage{enumitem}
\usepackage{multirow}
\usepackage{multicol}
\usepackage{stfloats}
\usepackage{tabularx} 
\usepackage{xpatch}
\usepackage{hyperref}
\usepackage{xcolor}
\usepackage{scalerel}
\usepackage{color}

\begin{document}
\title[Semi-sparsity for Triangular Mesh Denoising]
{Semi-sparsity on Piecewise Constant Function Spaces for Triangular Mesh Denoising}

\author[Junqing Huang,  Haihui Wang, Michael Ruzhansky \hfil \hfilneg] {Junqing Huang,  Haihui Wang and Michael Ruzhansky}  

\address{Junqing Huang \newline
Department of Mathematics: Analysis, Logic and Discrete Mathematics, Ghent University
Krijgslaan 281, Building S8 B 9000 Ghent Belgium}
\email{Junqing.Huang@UGent.be}

\address{Haihui Wang \newline
	School of Mathematical Sciences, BUAA University
	No. 37 Xueyuan Road, Haidian District, 100191 Beijing, China}
\email{whhmath@buaa.edu.cn}

\address{Michael Ruzhansky \newline
	Department of Mathematics: Analysis, Logic and Discrete Mathematics, Ghent University
	Krijgslaan 281, Building S8 B 9000 Ghent Belgium}
\email{Michael.Ruzhansky@UGent.be}

\subjclass[2000]{} 
\begin{abstract}

We present a semi-sparsity model for 3D triangular mesh denoising, which is motivated by the success of semi-sparsity regularization in image processing applications. We demonstrate that such a regularization model can be also applied for graphic processing and gives rise to the similar simultaneous-fitting results in preserving sharp features and piece-wise smoothing surfaces. Specifically, we first describe the piecewise constant function spaces associated with the differential operators on triangular meshes and then show how to extend the semi-sparsity model to meshes denoising. To verify its effectiveness, we  present an efficient iterative algorithm based on alternating direction method of multipliers (ADMM) technique and show the experimental results on synthetic and real scanning data against the state-of-the-arts both visually and quantitatively.

\end{abstract}

\maketitle \numberwithin{equation}{section}
\newtheorem{theorem}{Theorem}[section]
\newtheorem{corollary}[theorem]{Corollary}
\newtheorem{lemma}[theorem]{Lemma}
\newtheorem{remark}[theorem]{Remark}
\newtheorem{problem}[theorem]{Problem}
\newtheorem{example}[theorem]{Example}
\newtheorem{definition}[theorem]{Definition}
\allowdisplaybreaks

\section{Introduction}

Mesh denoising is a long-standing fundamental research topic in geometry processing. With the rapid development of 3D scanning devices, it has become increasingly popular and common to acquire and reconstruct meshes from the real world automatically. In many practical scenarios, it is inevidently for the acquired meshes to be contaminated by various noises because of local measurement errors in the scanning complex geometries and computational errors in reconstruction algorithms. As a result, it is highly expected to develop an effective denoising method to recover high-quality geometric structures from the corrupted acquiring data. However, it is a challenging problem because of the resemble high-frequency characteristics of geometric features and oscillating noises. 

In the literature, many techniques have been investigated to remove noise while preserving geometric features, including filtering methods\cite{fleishman2003bilateral,zhang2015guided,he2013mesh}, variational-based methods \cite{zhang2015variational, liu2020mesh} and higher-order variants\cite{liu2019triangulated, liu2021mesh}, and so on. For example, bilateral filter\cite{fleishman2003bilateral} and guided filter\cite{zhang2015guided} have been used for practical geometry processing due to the simplicity and ease of implementation, but they may cause over-smoothing effects around sharp edges limiting producing high-performance results. Variational-based methods have attracted great attention for mesh denoising, as they can well preserve sharp features while suppressing noise significantly.  Unfortunately, they may lead to stair-case artifacts in polynomial-smoothing surfaces. Recently, higher-order variational extensions such as total generalized variation (TGV) have been proposed to amend the potential stair-case artifacts, but they may still blur geometric features in case of strong noise and complex graphic features.

In general, existing methods have brought great progress in removing weak or small-scale edges and retaining strong or large-scale edges. However, it still has much space for improvement in removing local noise while reserving the local complex graphical features. In this paper, we also present a higher-order model for 3D triangular mesh denoising, which is based on semi-sparsity regularization to preserve sharp features and piece-wise smoothing surfaces. Specifically, we first describe the piecewise constant function spaces associated with the differential operators on triangular meshes and then show how to extend the semi-sparsity model to meshes denoising. To verify its effectiveness, we  present an efficient iterative algorithm based on alternating direction method of multipliers (ADMM) technique. The proposed method is also compared with the experimental results on synthetic and real scanning data against the state-of-the-arts both visually and quantitatively.

\section{Preliminaries}

In this section, we briefly introduce some notations and definitions of piecewise constant function spaces for the proposed semi-sparsity mesh denoisng model. The reader is also referred to ~\cite{botsch2010polygon, crane2018discrete, liu2019triangulated, liu2021mesh,  meyer2003discrete, zhang2015variational} for more details. 

\subsection{Notation}
Let $\mathcal{M}$ be a non-degenerate triangulated surface with vertices, edges, and triangles denoted as $v_{i(i=0,1,\cdots,I-1)}$, $e_{j(j=0,1,\cdots,E-1)}$ and $\tau_{k(k=0,1,\cdots,T-1)}$, respectively. We introduce the relative orientation of an edge $e$ to a triangle $\tau$, denoted by $s(e, \tau)$, where $v \prec e$ represents that $v$ is an endpoint of an edge $e$. Similarly, $e \prec \tau$ denotes that $e$ is an edge of a triangle $\tau ; v \prec \tau$ denotes that $v$ is a vertex of a triangle $\tau$.

We further introduce the relative orientation of an edge $e$ to a triangle $\tau$, which is denoted by $\operatorname{sgn}(e, \tau)$ as follows. We assume that all triangles have the counter-clockwise orientations and all edges are with randomly chosen fixed orientations. If an edge $e$ and a triangle $\tau$ have the same orientation, then $s(e, \tau) = 1$; otherwise, $s(e, \tau) = -1$. 

\begin{figure*}[!h]
	\begin{center}
		\subfigure[]
		{\includegraphics[width=0.245\textwidth]{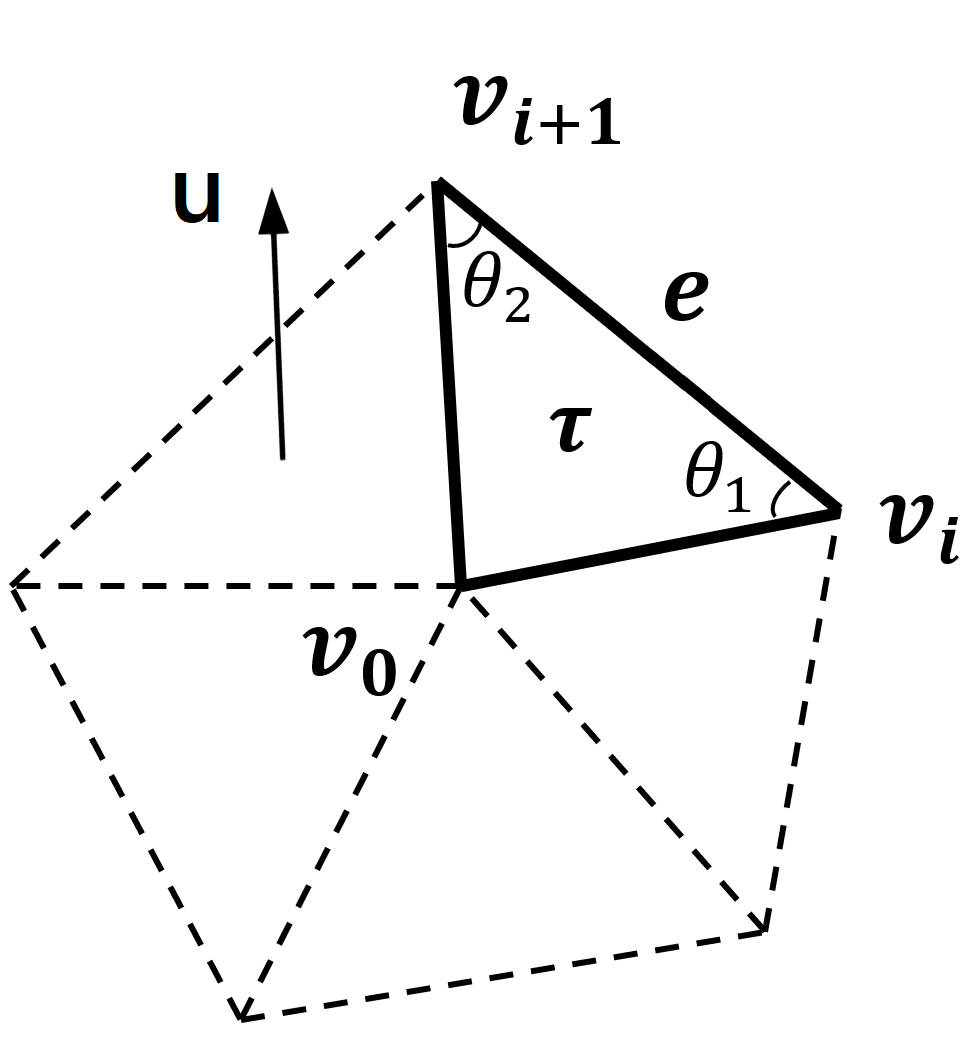}\label{Fig:fig1a}}
		\subfigure[]
		{\includegraphics[width=0.245\textwidth]{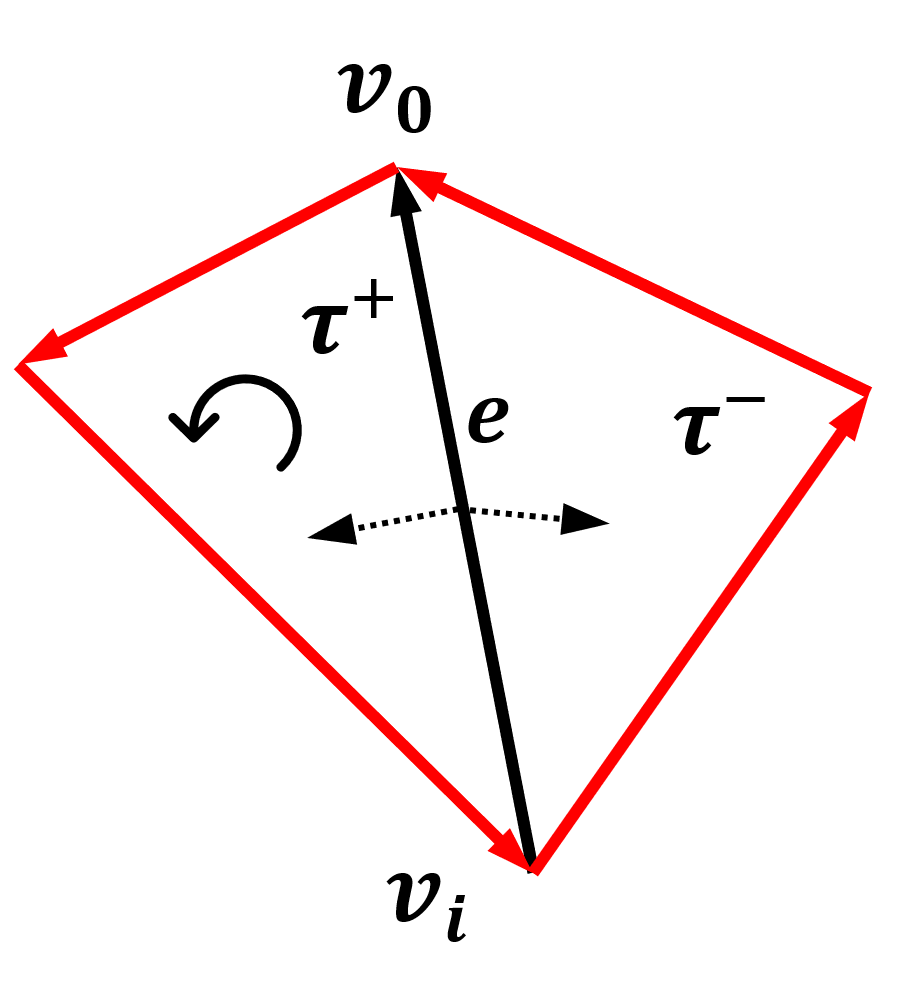}\label{Fig:fig1b}}
		\subfigure[]
		{\includegraphics[width=0.245\textwidth]{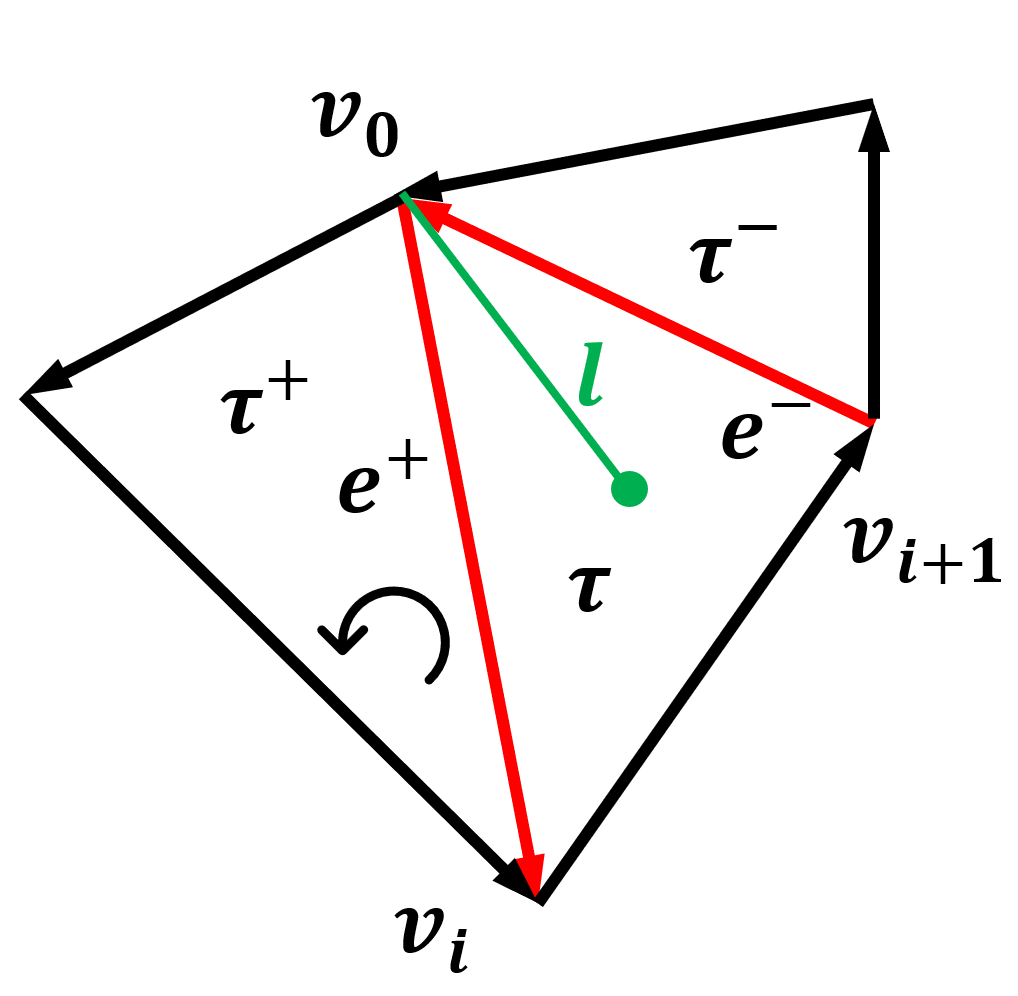}\label{Fig:fig1c}}
		\subfigure[]
		{\includegraphics[width=0.245\textwidth]{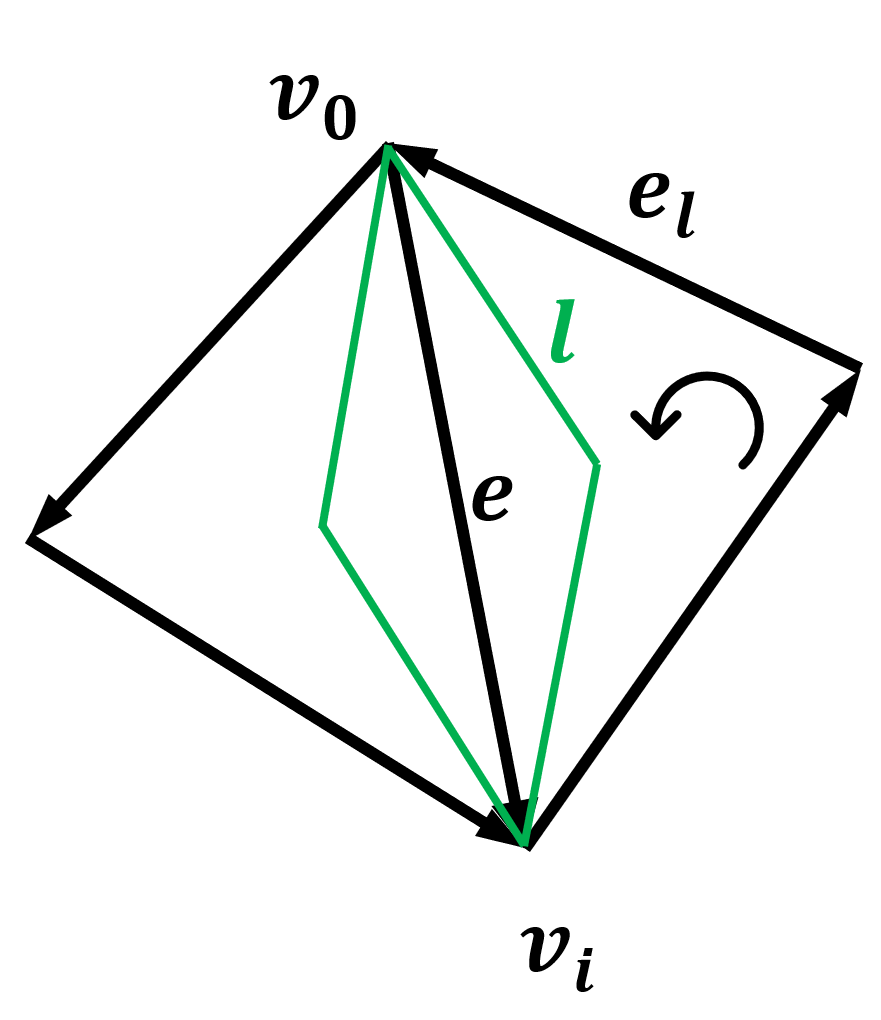}\label{Fig:fig1d}}
	\end{center}
	\caption{The illustration of discrete operators on meshes.}
	\label{Fig:fig1}
\end{figure*}

\subsection{Piece-wise linear Function Spaces and Differential Operators}

Piecewise constant function spaces have achieved great success in computer graphics applications\cite{liu2019triangulated, liu2021mesh, zhang2015variational}, because the spaces associated with differential operators form a basic easily-handled space to process graphic data such as triangular meshes. We introduce the spaces and show how to derive the assoiciated differential operators on triangulated meshes.

We define the space $U=\mathbb{R}^{\mathrm{T}}$, which is isomorphic to the piecewise constant function space on a triangulated mesh $\mathcal{M}$. Let $u=\left(u_0, \ldots, u_{\mathrm{T-1}}\right) \in U$ and $u_\tau$ 
be a vector restricted to triangle $\tau$, sometimes written as $\left.u\right|_\tau$ for convenience. For example, $u_{\tau}$ can be defined as the outward-facing normal vector restricted on triangle $\tau$, which, as shown in Fig. \ref{Fig:fig1a}, is perpendicular to plane defined by triangle $\tau$. According to \cite{liu2019triangulated, liu2021mesh}, the jump function of $u$ over an edge $e$  is defined as
\begin{equation}
[u]_e= 
\begin{cases}\sum_{e\prec\tau} \left. u\right|_r \operatorname{sgn}(e, \tau), & e \nsubseteq \partial \mathcal{M} \\ 0, & e \subseteq \partial \mathcal{M}
\end{cases}
\end{equation}
Here, the jump function $[u]_e$ can be illustrated in  Fig. \ref{Fig:fig1b}.

The space $U$ has the standard inner product and norm, 
\begin{equation}
	\left(u^1, u^2\right)_{U}=\sum_\tau \left.u^1\right|_\tau \left.u^2\right|_\tau s_{\mathrm{r}}, \quad\|u\|_{V}=\sqrt{(u, u)_{U}},
\end{equation}
where $u^1, u^2$ and $u \in V$, and $s_\tau$ is the area of triangle $\tau$.

Let the edge function space $V=\mathbb{R}^{\mathrm{E}}$ and $v_e\in V $ (or $\left.v\right|_e$) be a vector restricted to edge $e$, it is then natural to define the first-order differential operator $\mathcal{D}:U \mapsto V$ on $\mathcal{M}$ as
\begin{equation}
		\begin{aligned}
 \quad \left.\mathcal{D} u\right|_e=[u]_e, \quad \forall e, \; \text{for} \; u \in V.
\end{aligned}
 \label{Eq:eq2-3} 
\end{equation}

By definition, the space $V$ is also equipped with the inner product and norm:
\begin{equation}
\left(v^1, v^2\right)_{V}=\sum_e \left.v^1\right|_e \left.v^2\right|_e \operatorname{len}(e), \quad\|v\|_{V}=\sqrt{(v, v)_{V}},
\end{equation}
where $v^1, v^2, v \in V$, and $len(e)$ is the length of the edge $e$. 

As explained in \cite{liu2019triangulated, liu2021mesh}, the adjoint operator of $\mathcal{D}$, that is, $\mathcal{D}^\ast: U \mapsto V$ given by

\begin{equation}
		\begin{aligned}
\left.(\mathcal{D}^\ast v)\right|_\tau=-\frac{1}{s_\tau} \sum_{\substack{e \prec \tau \\ e \not \subset \partial \mathcal{M}}} \left.v\right|_e \operatorname{sgn}(e, \tau) \operatorname{len}(e), \; \forall \tau.
\end{aligned}
\label{Eq:eq2-5} 
\end{equation}

Eq. \ref{Eq:eq2-3} and Eq. \ref{Eq:eq2-5} define the first-order differential operator and the associated adjoint operator on the triangled mesh $\mathcal{M}$. 

By analogy, it is easy to define the higher-order differential operators. Let $l$ be the line connecting the barycenter and one vertex of the triangle $\tau$. As depicted in Fig. \ref{Fig:fig1c}, the two edges ${e}^{+}$ and ${e}^{-}$ share the common vertex of $l$, and the two triangles sharing edges $e^{+}$ and $e^{-}$ are denoted as ${\tau}^{+}$ and ${\tau}^{-}$, respectively. We then define the jump difference over the line $l$ as $[[u]]_{l, \tau}$  (or, $\left[\left[u\right]\right]_l$),
\begin{equation}
	\begin{aligned}
		\left[\left[u\right]\right]_{l,\tau}
		&= \left[u\right]_{e^+}sgn(e^+, {\tau}^{+})  \!+ \! \left[u\right]_{e^-}sgn(e^-,{\tau}^{-})\\
		&= (u_{\tau^{+}} \!-\! u_{\tau})-(u_{\tau} \!-\! u_{\tau^{-}}) \\
		&= u_{\tau^{+}} \!-\! 2u_{\tau}\!+\! u_{\tau^{-}}.
	\end{aligned}
	\label{Eq:eq2-6}  
\end{equation}
It is clear that Eq. \ref{Eq:eq2-6} can be viewed as a second order difference operator with respect to $u$. For any $u \in U$ with the Neumann boundary condition, we actually have
\begin{equation}
[[u]]_l=\left\{
\begin{aligned}
	u_{\tau+}-2 u_\tau+u_{\tau^{-}}, &\quad e^{+} \text {or } e^{-} \not \subset \partial \mathcal{M} \\
	0, &\quad e^{+} \text {or } e^{-} \subset \partial \mathcal{M}
\end{aligned}
\right.
\label{f2} 
\end{equation}
We can see that $\left[\left[u\right]\right]_l$ is invariant under the choice of orientation of edges.

In the discrete case, for each triangle $\tau$, there are three first-order differences over the edges along three different directions. Thus, we have the gradient operator in $\tau$ as,
\begin{equation}
	\left.\nabla u\right|_\tau=\left(\left.\mathcal{D}_{\mathcal{M}} u\right|_{e_{1, \tau}},\left.\mathcal{D}_{\mathcal{M}} u\right|_{e_{2, \tau}},\left.\mathcal{D}_{\mathcal{M}} u\right|_{e_{3, \tau}}\right),
\end{equation}
where $e_{i, \tau} \prec \tau, i=1,2,3$.  We may write the discrete gradient as $\nabla u=\left(\partial_1 u, \partial_2 u, \partial_3 u\right)$ for convenience. 

Similarity, it is natural to denote the second-order gradient with respect to $u$ restricted to $\tau$ as,
$\nabla^2: U \rightarrow W, u \mapsto \nabla^2 u,\left.\quad \nabla^2 u\right|_\tau=\left([[u]]_{l_{0, \tau}},[[u]]_{l_{1, \tau}},[[u]]_{l_{2, \tau}}\right), \forall \tau$, for $u \in V$, where $W=\mathbb{R}^{\mathrm{T}} \times \mathbb{R}^{\mathrm{T}} \times \mathbb{R}^{\mathrm{T}}$.
For convenience, we may also write it in the form
\begin{equation}
\left.\nabla^2 u\right|_\tau=\left(\begin{array}{lll}
	\partial_1 \partial_1 u & \partial_1 \partial_2 u & \partial_1 \partial_3 u \\
	\partial_2 \partial_1 u & \partial_2 \partial_2 u & \partial_2 \partial_3 u \\
	\partial_3 \partial_1 u & \partial_3 \partial_2 u & \partial_3 \partial_3 u
\end{array}\right),
\end{equation}
where the diagonal entries $\partial_i \partial_i u, i=\{1,2,3\}$ are the second-order directional derivatives in the same direction, while the off-diagonal entries $\partial_i \partial_j u, i \neq j$ are the second-order directional derivatives in two different directions. It is also possible to define the higher-order differential operator for the edge $e$ as shown in Fig. \ref{Fig:fig1d}

\subsection{Piecewise Constant Function Spaces}

To handle vectorial data, we extend the above concepts to vectorial cases with the definitions of spaces $\mathbf{U}, \mathbf{V}$ and $\mathbf{W}$ as follows:
\begin{equation}
\mathbf{U}=\underbrace{U \times \cdots \times U}_{\mathfrak{N}},
\mathbf{V}=\underbrace{V \times \cdots \times V}_{\mathfrak{N}}, 
\mathbf{W}=\underbrace{W \times \cdots \times W}_{\mathfrak{N}},
\end{equation}
for $\mathfrak{N}$-channel data. The inner products and norms in $\mathbf{U}, \mathbf{V}$ and $\mathbf{W}$ are as follows:
\begin{equation}
\begin{aligned}
	\left(\mathbf{u}^1, \mathbf{u}^2\right)_{\mathbf{U}} & =\sum_{1 \leq i \leq \mathfrak{N}}\left(u_i^1, u_i^2\right)_{U}, \quad\|\mathbf{u}\|_{\mathbf{V}}=\sqrt{(\mathbf{u}, \mathbf{u})_{\mathbf{U}}}, \mathbf{u}^1, \mathbf{u}^2, \mathbf{u} \in \mathbf{U}, \\
	\left(\mathbf{v}^1, \mathbf{v}^2\right)_{\mathbf{V}} & =\sum_{1 \leq i \leq \mathfrak{N}}\left(v_i^1, v_i^2\right)_{V}, \quad\|\mathbf{v}\|_{\mathbf{V}}=\sqrt{(\mathbf{u}, \mathbf{v})_{\mathbf{V}}}, \mathbf{v}^1, \mathbf{v}^2, \mathbf{v} \in \mathbf{V}, \\
	\left(\mathbf{w}^1, \mathbf{w}^2\right)_{\mathbf{W}} & =\sum_{1 \leq i \leq \mathfrak{N}}\left(w_i^1, w_i^2\right)_{W}, \quad\|\mathbf{w}\|_{\mathbf{W}}=\sqrt{(\mathbf{w}, \mathbf{w})_{\mathbf{W}}}, \mathbf{w}^1, \mathbf{w}^2, \mathbf{w} \in \mathbf{W} .
\end{aligned}
\end{equation}
We mention that $\nabla \mathbf{u}, \nabla^2 \mathbf{u}$ and their adjoint operators can be computed channel by channel.

\section {Semi-sparsity Regularization for Mesh Denoising} 

Similar to many filtering methods that are firstly proposed for image processing and then applied in graphic processing~\cite{fleishman2003bilateral, he2013mesh, zhang2015guided}, it is straightforward to extend the semi-sparse model~\cite{huang2023semi} to 3D geometry, because the 3D meshes suffer from the similar piece-wise constant and smoothing surfaces with discontinuous boundaries. 
The semi-sparsity model is a higher-order case of sparse regularization that enables us to smooth 3D meshes without causing stair-case artifacts.

\subsection{Problem Formulation}

According to \cite{huang2023semi}, the semi-sparsity prior knowledge of signal is suggested to be formulated into a higher-order $L_0$ regularization model in the context of optimization-based framework, which has a general following form,
\begin{equation}
	\begin{aligned}
		\mathop{\min}_{u} {\frac{\beta}{2}{\left\Vert {u}-{f}\right\Vert }_2^2}+\alpha_1 \sum_{k=1}^{n-1}  {\left\Vert {\nabla}^{k}u-{\nabla}^{k} f \right\Vert }_p^p+\alpha_2{\left\Vert {\nabla}^{n} u \right\Vert }_0
	\end{aligned}
	\label{Eq:eq3-1} 
\end{equation}
where $u$ and $f$ are the target output and observation signals (images, 3D meshes, etc.), respectively. $\beta$,  $\alpha_1$ and $\alpha_2$ weigh the balance of three terms. The first term in Eq. \ref{Eq:eq3-1} is data fidelity to in a sense of least square minimization. The second term measures the $L_p(p \ge 1)$- norm similarity of higher-order gradients ${\nabla}^{k}u$ and ${\nabla}^{k}f$ in consideration of the piece-wise polynomial surfaces. The third term  ${\left\Vert {\nabla}^{n} u \right\Vert }_0$ favors the highest-order gradient ${ {\nabla}^{n} u}$ to be fully sparse. The idea of Eq. \ref{Eq:eq3-1} is straightforward, that is, a sparse-induced $L_0$-norm constraint is only imposed on the highest order $n\!-\!th$ gradient domain, as the ones with the orders less than $n$ are not fully sparse but also have a small error $L_p$ space. 

The above claims are also valid for the piece-wise smoothing surfaces of 3D meshes. Given a noisy mesh $\mathcal{M}$ with the normal field denoted as $\mathbf{N}_0$, we have the semi-sparse regularization for normal filter as the following problem, 
\begin{equation}
	\begin{aligned}
		\mathop{\min}_{u} {\frac{\beta}{2}{\left\Vert \mathbf{N}-\mathbf{N}_0 \right\Vert }_{\mathbf{U}}^2}+\alpha_1 {\left\Vert {\nabla}\mathbf{N} - {\nabla}\mathbf{N}_0\right\Vert }_{\mathbf{V}}+\alpha_2{\left\Vert {\nabla}^{2}\mathbf{N} \right\Vert }_{\mathbf{W}}^0.
	\end{aligned}
	\label{Eq:eq3-2} 
\end{equation}
Here, as indicated in \cite{huang2023semi}, we set highest order $n=2$ and $p=1$ in Eq. \ref{Eq:eq3-1} for the sake of simplicity and computational efficiency.

\subsection{The Efficient ADMM Solver}

Due to the non-smooth and non-convex objective function of Eq.\ref{Eq:eq3-2}, a direct solution is not available. Instead, we propose to solve the problem based on an alternating direction method of multipliers (ADMM), which has achieved great success in solving the related problems \cite{liu2019triangulated, liu2021mesh}.

By introducing the new variables $\mathrm{P}$, and $\mathrm{Q}$, we then reformulate Eq. \ref{Eq:eq3-2} as a constrained optimization problem with the following form,
\begin{equation}
	\begin{aligned}
	\min_{\mathbf{N}, \mathbf{P}, \mathbf{Q}} 	&\left\{
	\frac{\beta}{2}\left\|\mathbf{N}-\mathbf{N}_0\right\|_{\mathbf{U}}^2
	+\alpha_1\left\|\mathbf{P}\right\|_{\mathbf{V}}
	+\alpha_2\left\|\mathbf{Q}\right\|_{\mathbf{W}}^0+ \Psi (\mathbf{N})
	\right\} , \\
	\text {s.t.} \quad & \qquad \mathbf{P}={\nabla}\mathbf{N} - {\nabla}\mathbf{N}_0, \quad \mathbf{Q}={\nabla}^{2}\mathbf{N},
	\end{aligned}
	\label{Eq:eq3-3} 
\end{equation}
where
\begin{equation}
	\Psi(\mathbf{N})=\left\{
\begin{aligned}
	0, &\qquad\left\|\mathbf{N}_{\tau} \right\| = 1, \forall \tau,\\
	+\infty, &\qquad \quad \text{otherwise}.
\end{aligned}
	\right.
	\label{Eq:eq3-4}
\end{equation}

Accordingly, we introduce the augmented Lagrangian function of the above constrained optimization problem,
\begin{equation}
	\begin{aligned}
	\mathcal{L}\left(\mathbf{N}, \mathbf{P}, \mathbf{Q}, \lambda_{\mathbf{P}}, \lambda_{\mathbf{Q}}\right)& =
	\frac{\beta}{2}\left\|\mathbf{N}-\mathbf{N}_0\right\|_{\mathbf{U}}^2
	+\alpha_1\left\|\mathbf{P}\right\|_{\mathbf{V}}
	+\alpha_2\left\|\mathbf{Q}\right\|_{\mathbf{W}}^0+ \Psi (\mathbf{N}) \\
	& +\langle\lambda_{\mathbf{P}}, \left({\nabla}\mathbf{N}-{\nabla}\mathbf{N}_0\right)-\mathbf{P}\rangle_{\mathbf{V}}+\frac{\rho_1}{2}\left\|\left(\mathbf{N}-\mathbf{N}_0\right)-\mathbf{P}\right\|_{\mathbf{V}}^2 \\
	& +\langle\lambda_{\mathbf{Q}}, {\nabla}^{2}\mathbf{N}-\mathbf{Q}\rangle_{\mathbf{W}}+\frac{\rho_2}{2}\left\|{\nabla}^{2}\mathbf{N}-\mathbf{Q}\right\|_{\mathbf{W}}^2,
	\end{aligned}
\label{Eq:eq3-5} 
\end{equation}
where $\lambda_{\mathbf{P}}$, and $\lambda_{\mathbf{Q}}$ are Lagrange multipliers, $\rho_1$ and $\rho_2$ are positive penalty weights. The variable-splitting technique is then applied to iteratively update the variables in an alternative way, giving the the following  sub-problems:

\subsubsection{The $\mathbf{N}$-subproblem:}
\begin{equation}
\begin{aligned}
	\min_{\mathbf{N}}  &\frac{\beta}{2} \|\mathbf{N}-\mathbf{N}_0\|_{\mathbf{U}}^2 
	 +\frac{\rho_1}{2}\left\|\left( {\nabla}\mathbf{N}-{\nabla}\mathbf{N}_0\right)-\mathbf{P}+\frac{\lambda_{\mathbf{P}}}{\rho_1}\right\|_{\mathbf{V}}^2\\
	 +&\frac{\rho_2}{2}\left\| {\nabla}^2\mathbf{N}-\mathbf{Q}+\frac{\lambda_{\mathbf{Q}}}{\rho_2}\right\|_{\mathbf{V}}^2+\Psi(\mathbf{N}).
\end{aligned}
\label{Eq:eq3-6}
\end{equation}

It is clear that Eq. \ref{Eq:eq3-6} is a quadratic optimization problem with the unit normal constraints. As suggested in \cite{liu2021mesh}, an approximation strategy is employed to solve this problem. We first solve the quadratic program and then project the solution $\mathbf{N}$ onto a unit sphere. Specifically,  the corresponding Euler-Lagrange equation based on the first-order optimal conditions has the form
\begin{equation}
\begin{aligned}
\beta \left(\mathbf{N} -\mathbf{N}_0\right) -\rho_1{\nabla}^{\ast}\left(  {\nabla}\mathbf{N}-{\nabla}\mathbf{N}_0-\mathbf{P}+\frac{\lambda_{\mathbf{P}}}{\rho_1}\right)+\rho_2\left({\nabla}^2\right)^{\ast}\left( {\nabla}^2\mathbf{N}-\mathbf{Q}+\frac{\lambda_{\mathbf{Q}}}{\rho_2}\right)=0.
\end{aligned}
\label{Eq:eq3-7}
\end{equation}
where ${\nabla}^{\ast}$ and $\left({\nabla}^2\right)^{\ast}$ are the adjoint operators of the first-order and second-order differential operators, respectively. The above equation is a sparse and positive semi-definite linear system, which can be solved by efficient sparse linear solvers.

\subsubsection{The $\mathbf{P}$-subproblem:}
\begin{equation}
\begin{aligned}
\min_{\mathbf{P}} \alpha_1\left\|\mathbf{P}\right\|_{\mathbf{V}}+\frac{\rho_1}{2}\left\|\mathbf{P}-\left( {\nabla}\mathbf{N}+{\nabla}\mathbf{N}_0\right)+\frac{\lambda_{\mathbf{P}}}{\rho_1}\right\|_{\mathbf{V}}^2, 
\end{aligned}
\label{Eq:eq3-8}
\end{equation}
where \ref{Eq:eq3-8} is a classical Lasso problem and can be solved efficiently by splitting  each variable $\mathbf{P}_e$ independently, where the $\mathbf{P}_e$ has a closed form solution 
\begin{equation}
\begin{aligned}
\mathbf{P}_e=\mathcal{S}\left(\left( {\nabla}\mathbf{N}+{\nabla}\mathbf{N}_0\right)-\frac{\lambda_{\mathbf{P}}}{\rho_1}, \frac{\alpha_1}{\rho_1}\right),
\end{aligned}
\label{Eq:eq3-9}
\end{equation}
with the soft shrinkage operator $\mathcal{S}(x,T)$ defined as:
$$
\mathcal{S}(x,T)= \operatorname{sign}(x) \max\left(0, \left\|x\right\|-T \right) .
$$

\subsubsection{The $\mathbf{Q}$-subproblem:}
\begin{equation}
\begin{aligned}
\min_{\mathbf{Q}} \alpha_2\left\|\mathbf{Q}\right\|_{\mathbf{W}}^0+\frac{\rho_2}{2}\left\|\mathbf{Q}- {\nabla}^2\mathbf{N}+\frac{\lambda_{\mathbf{Q}}}{\rho_2}\right\|_{\mathbf{V}}^2.
\end{aligned}
\label{Eq:eq3-10}
\end{equation}

The $\mathbf{Q}$-subproblem \ref {Eq:eq3-10} is a $L_0$ norm minimization problem, which has a similar separable property as \ref {Eq:eq3-8} and each variable $\bar{Q}_l$ is given by the formula
\begin{equation}
	\begin{aligned}
\mathbf{Q}_l=\mathcal{H}\left({\nabla}^2\mathbf{N}-\frac{\lambda_{\mathbf{Q}}}{\rho_2}, \frac{\alpha_2}{\rho_2}\right),
\end{aligned}
\label{Eq:eq3-11}
\end{equation}
with the hard-threshold operator defined as:
$$
	\mathcal{H}(x,T)=\left\{
	\begin{aligned}
		0, &\qquad\left\|x\right\|\le T,\\
		x, &\qquad\text{otherwise}.
	\end{aligned}
	\right.
$$

Finally, the Lagrange multipliers $\lambda_{\mathbf{P}}$ and $\lambda_{\mathbf{Q}}$ are updated in the form,
\begin{equation}
	\begin{aligned}
		\lambda_{\mathbf{P}} = &\lambda_{\mathbf{P}} + \rho_1\left( \left({\nabla}\mathbf{N}-{\nabla}\mathbf{N}_0\right)-\mathbf{P} \right),\\
		\lambda_{\mathbf{Q}} = &\lambda_{\mathbf{Q}} + \rho_2\left( {\nabla}^2\mathbf{N}-\mathbf{Q} \right).
	\end{aligned}
	\label{Eq:eq3-11}
\end{equation}

In summary, the semi-sparsity model in Eq. \ref{Eq:eq3-2} for normal filter is achieved by solving the ADMM subproblems and Lagrange multipliers iteratively. The procedure terminates when one of the stopping criteria is met. The scheme is also verified by the numeral results in the next section.

\section{EXPERIMENTAl RESULTS}

\begin{figure*}[!t]
	\begin{center}
		\subfigure[Noisy]
		{\includegraphics[width=0.2\textwidth]{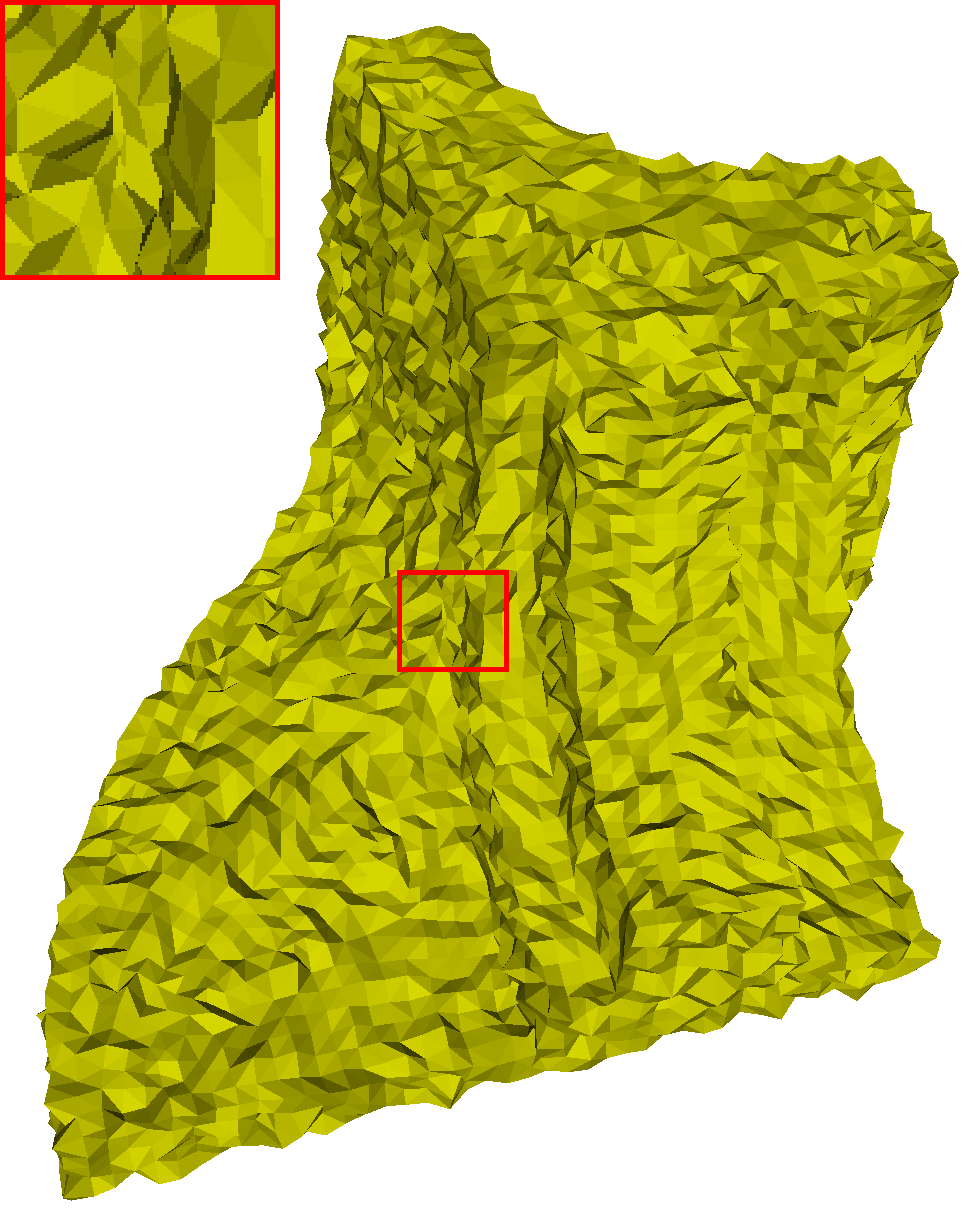}}
		\subfigure[BF\cite{fleishman2003bilateral}]
		{\includegraphics[width=0.2\textwidth]{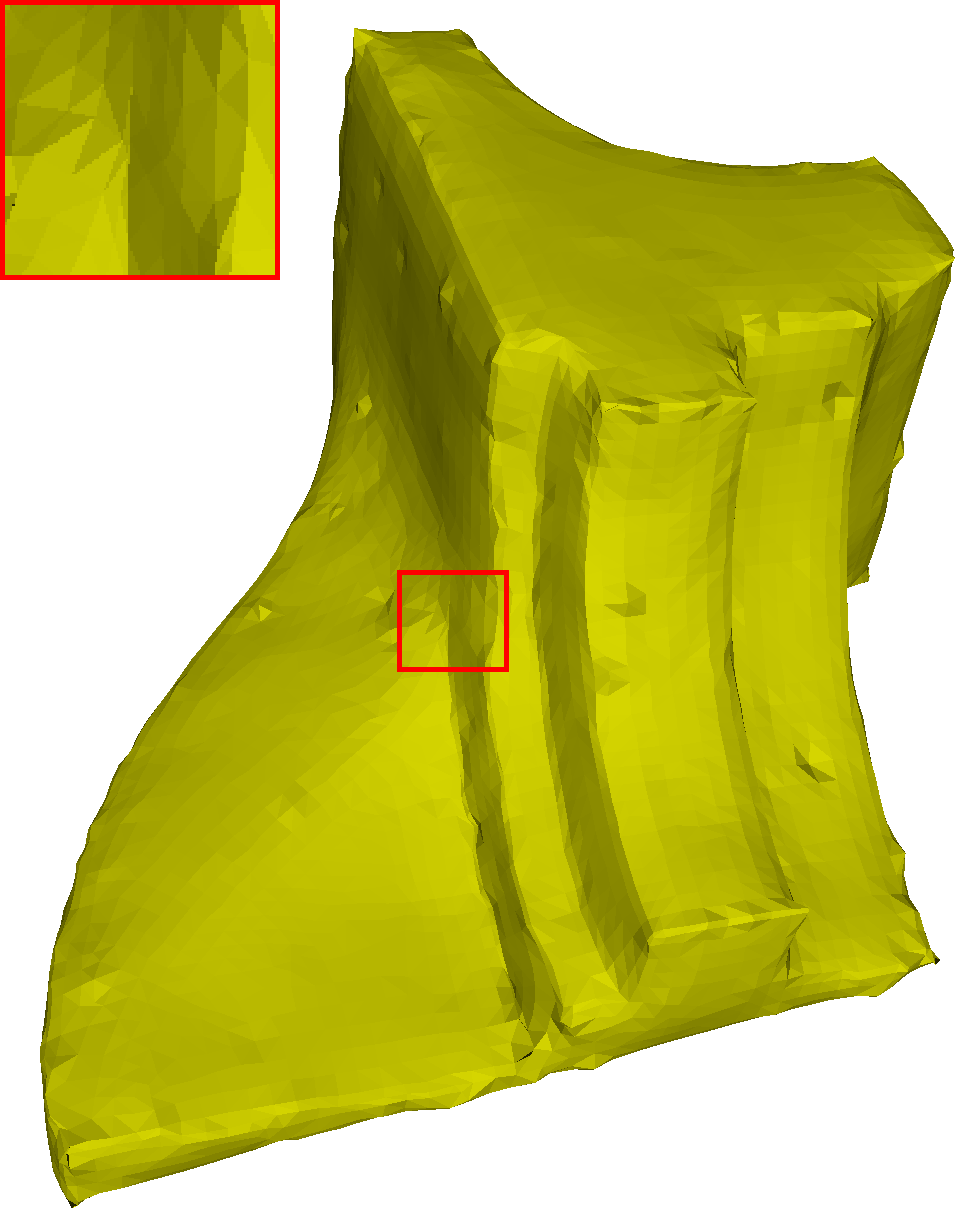}}
		\subfigure[GNF\cite{zhang2015guided}]
		{\includegraphics[width=0.2\textwidth]{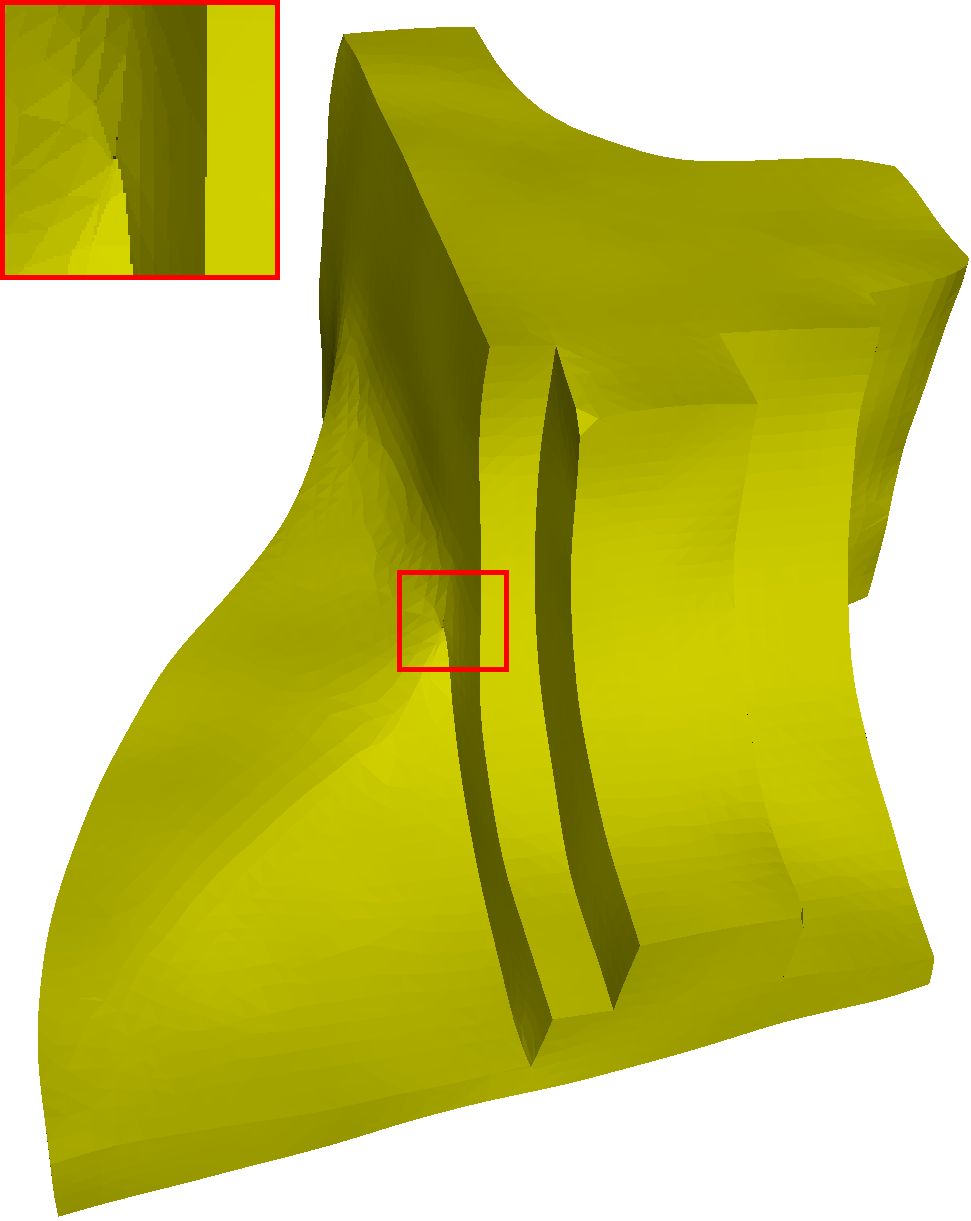}}
		\subfigure[CNR\cite{wang2016mesh}]
		{\includegraphics[width=0.2\textwidth]{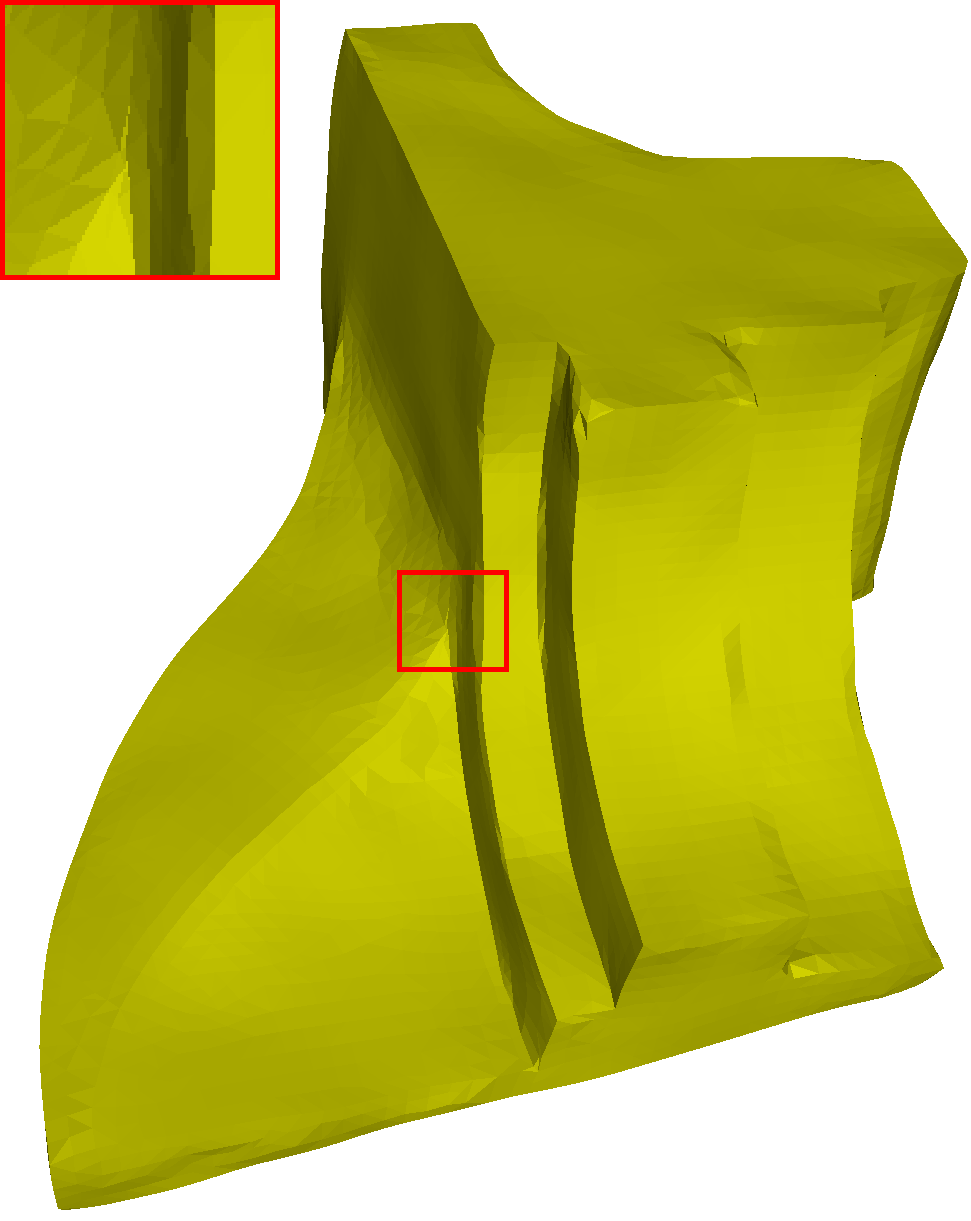}}
		
		\subfigure[$L_0$\cite{he2013mesh}]
		{\includegraphics[width=0.2\textwidth]{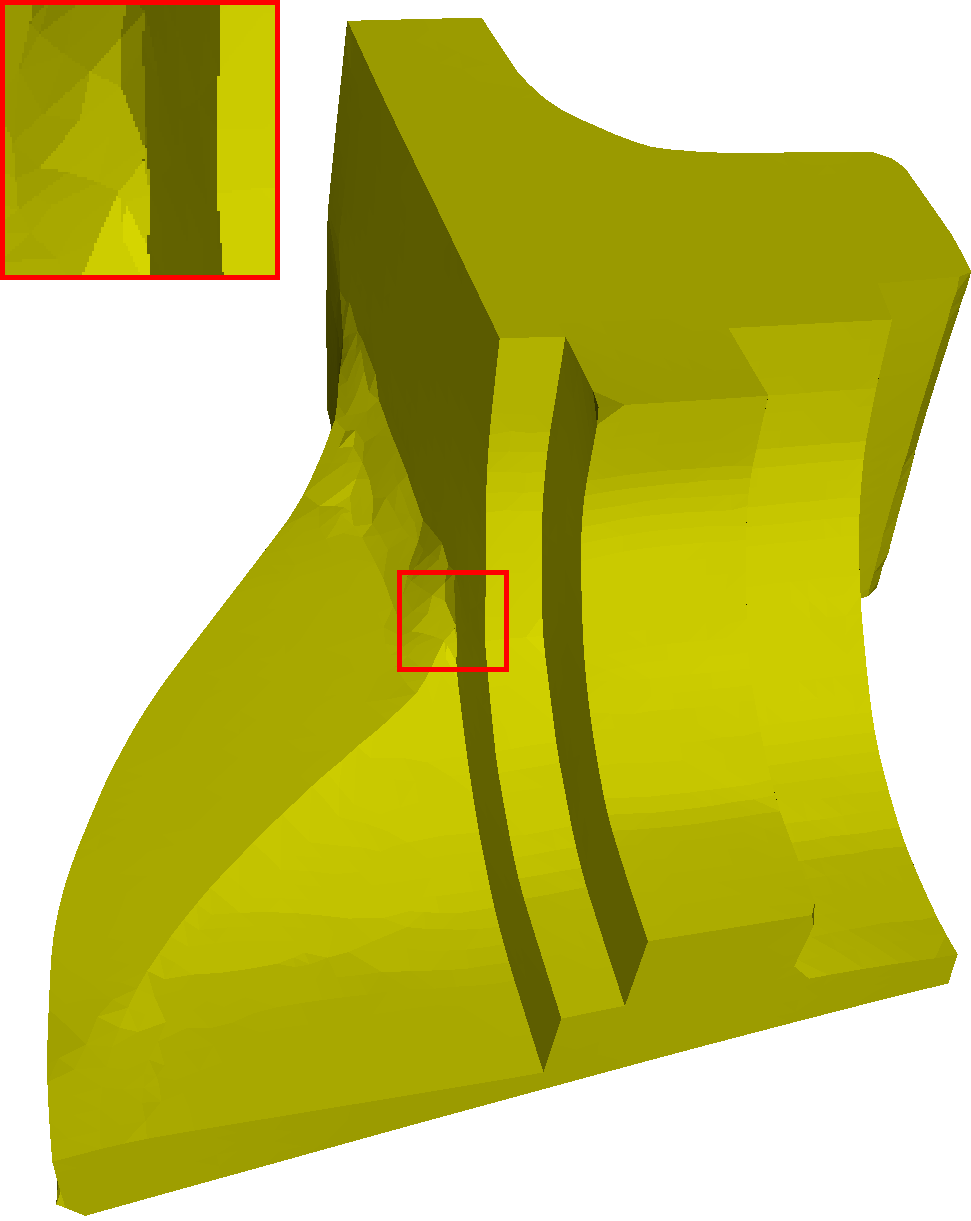}}
		\subfigure[TGV\cite{liu2021mesh}]
		{\includegraphics[width=0.2\textwidth]{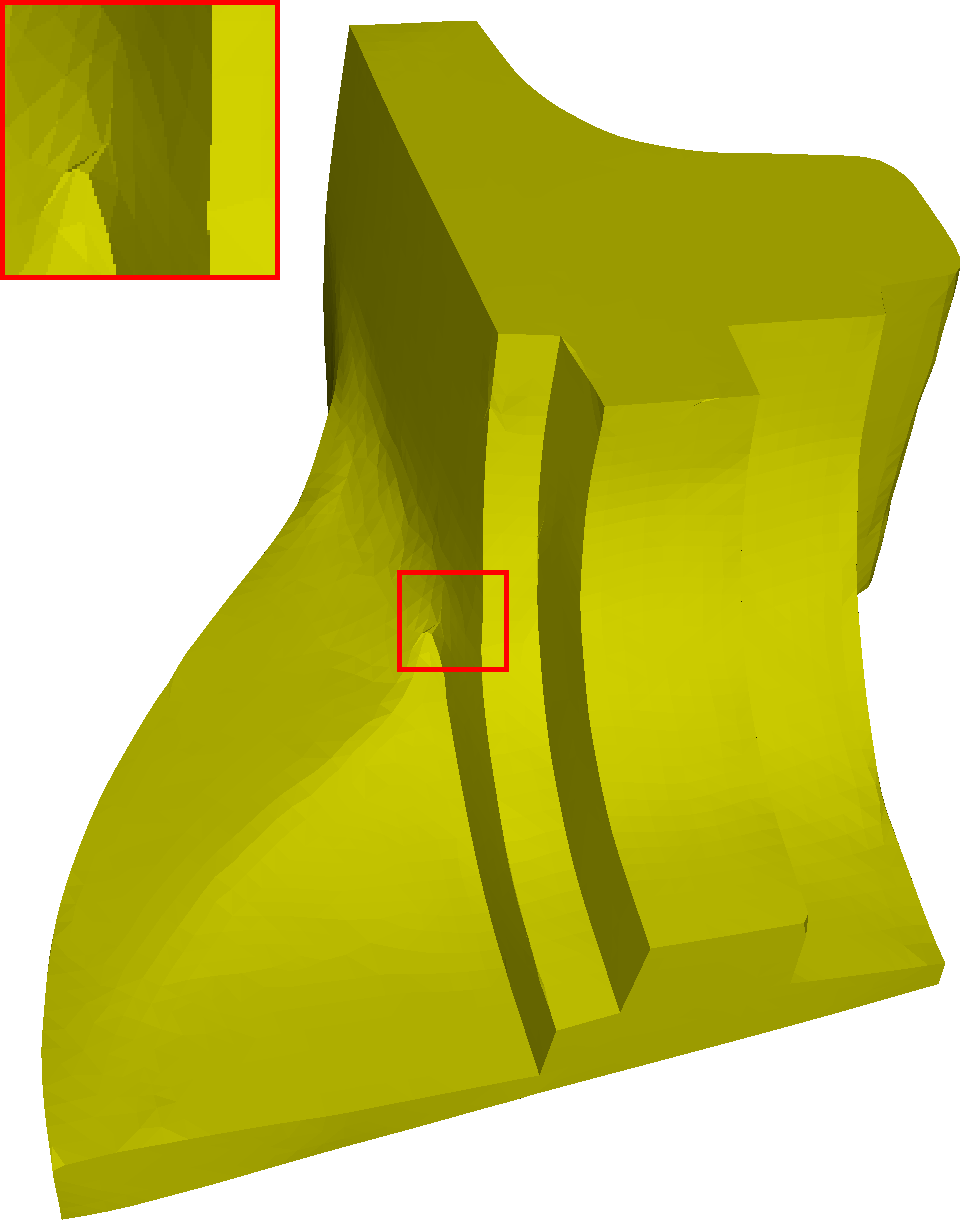}}
		\subfigure[Ours]
		{\includegraphics[width=0.2\textwidth]{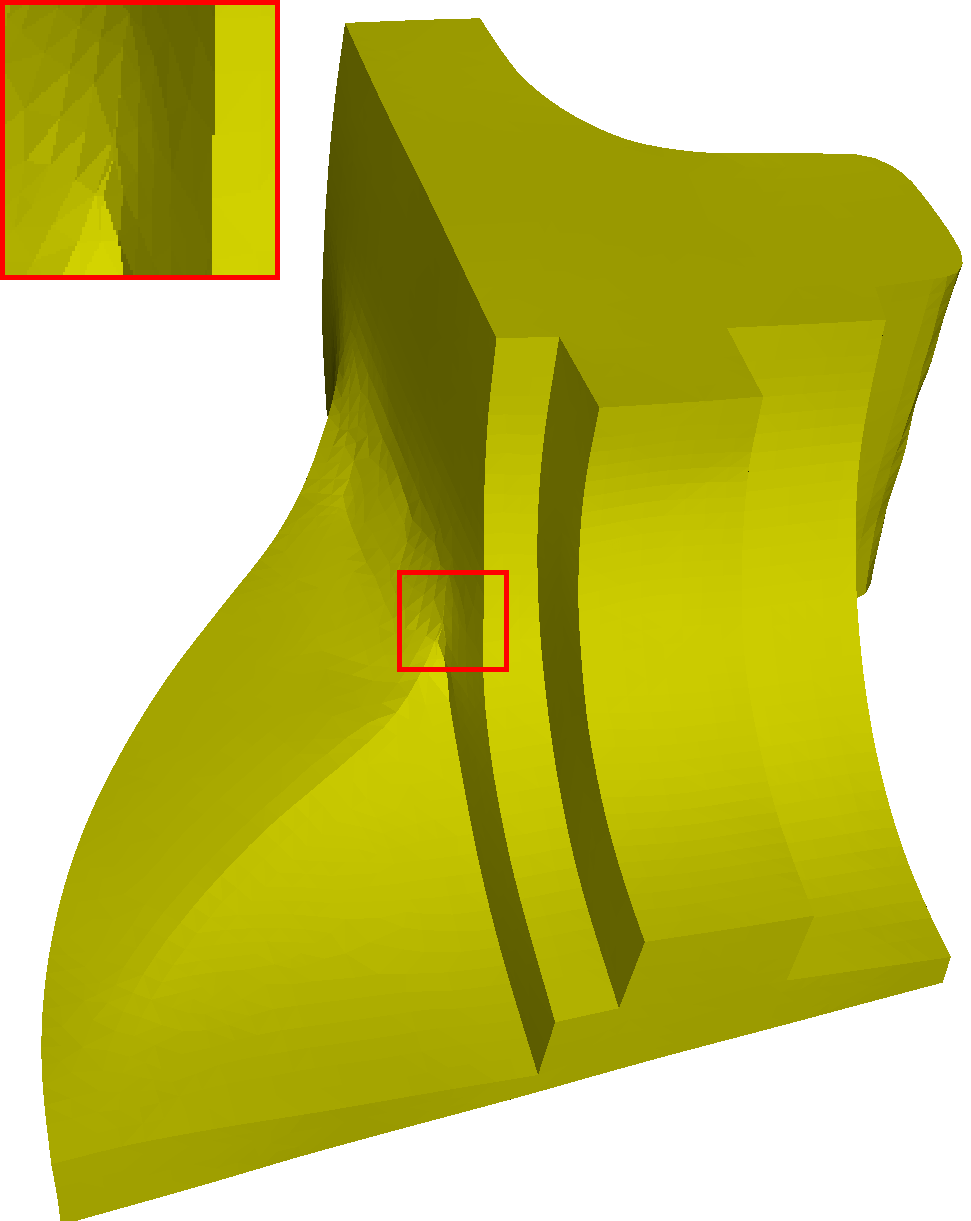}}
		\subfigure[GT]
		{\includegraphics[width=0.2\textwidth]{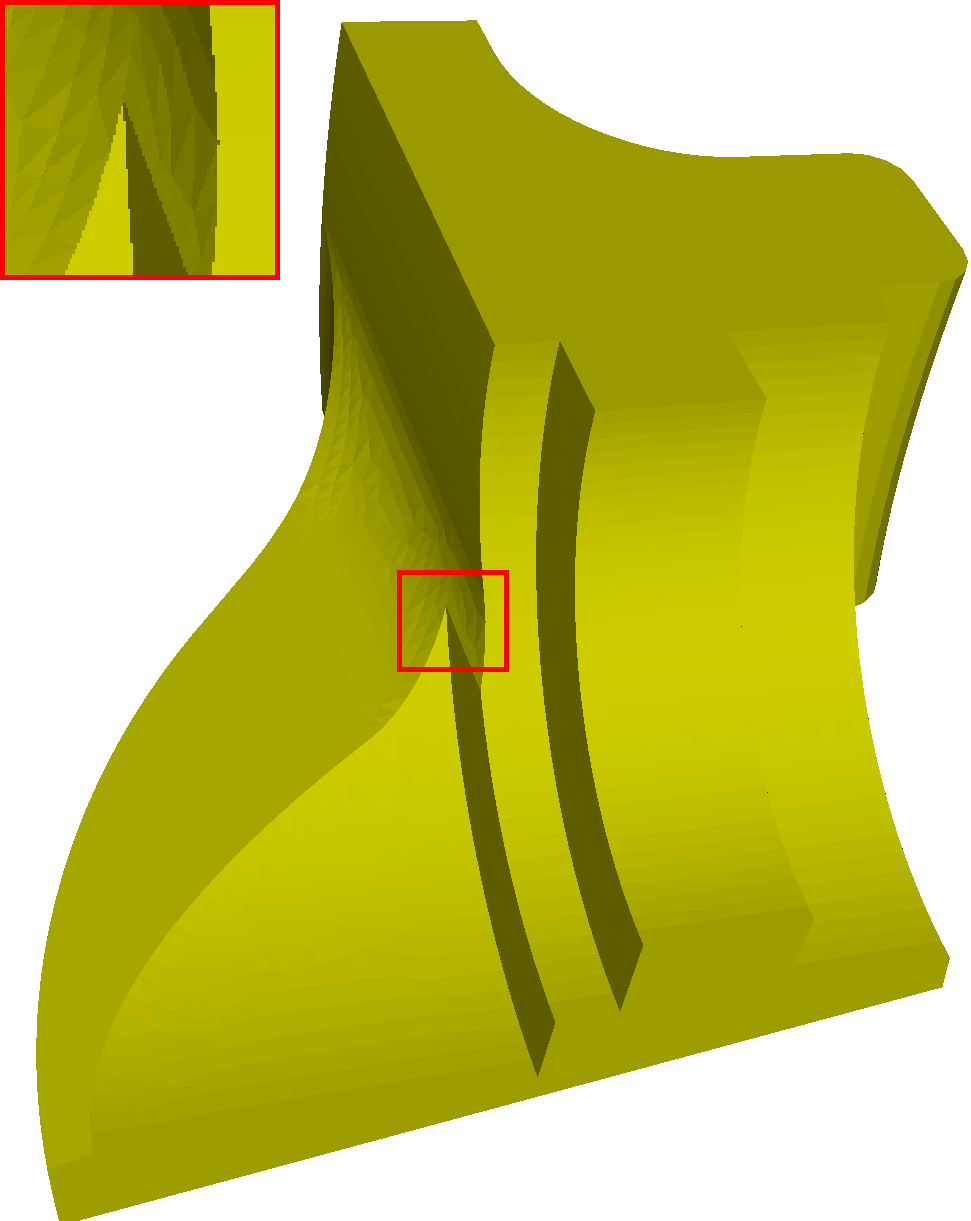}}
	\end{center}
	\caption{Mesh denoising, surface corrupted by Gaussian noise in random directions with standard deviation $\sigma = 0.3\bar{L}$ ($\bar{L}$ is the average length of edges). (a) Noisy input, (b) Bilateral filter (BF)\cite{fleishman2003bilateral}, (c) Guided normal filter (GNF)\cite{zhang2015guided}, (d) Cascaded normal regression (CNR)\cite{wang2016mesh}, (e) $L_0$ minimization\cite{he2013mesh}, (f) TGV regularization\cite{liu2019triangulated}, (g) Our result and (h) Ground Truth (GT).}
	\label{Fig:fig2}
\end{figure*} %

We have interpreted the definitions of the differential operators on triangular meshes in the previous section. Once they are computed, it is easy to substitute them into the semi-sparsity model and solve it based on the ADMM algorithm accordingly. In order to further illustrate the proposed semi-sparsity model, we here compare it with the existing mesh denoising methods, including the feature-aware mesh filter ~\cite{sun2007fast},bilateral filter (BF)\cite{fleishman2003bilateral}, guided normal filter (GNF)\cite{zhang2015guided}, cascaded normal regression (CNR)\cite{wang2016mesh}, $L_0$ minimization\cite{he2013mesh}, and high-order TGV regularization\cite{liu2021mesh}. We carefully tune the parameters of each competing methods so that satisfactory results are produced.  Our Matlab implementation runs on the PC with Intel Core2 Duo CPU 2.13G and $32 \mathrm{~GB}$ RAM. 

\begin{figure*}[!h]
	\begin{center}
		{\includegraphics[width=0.2\textwidth]{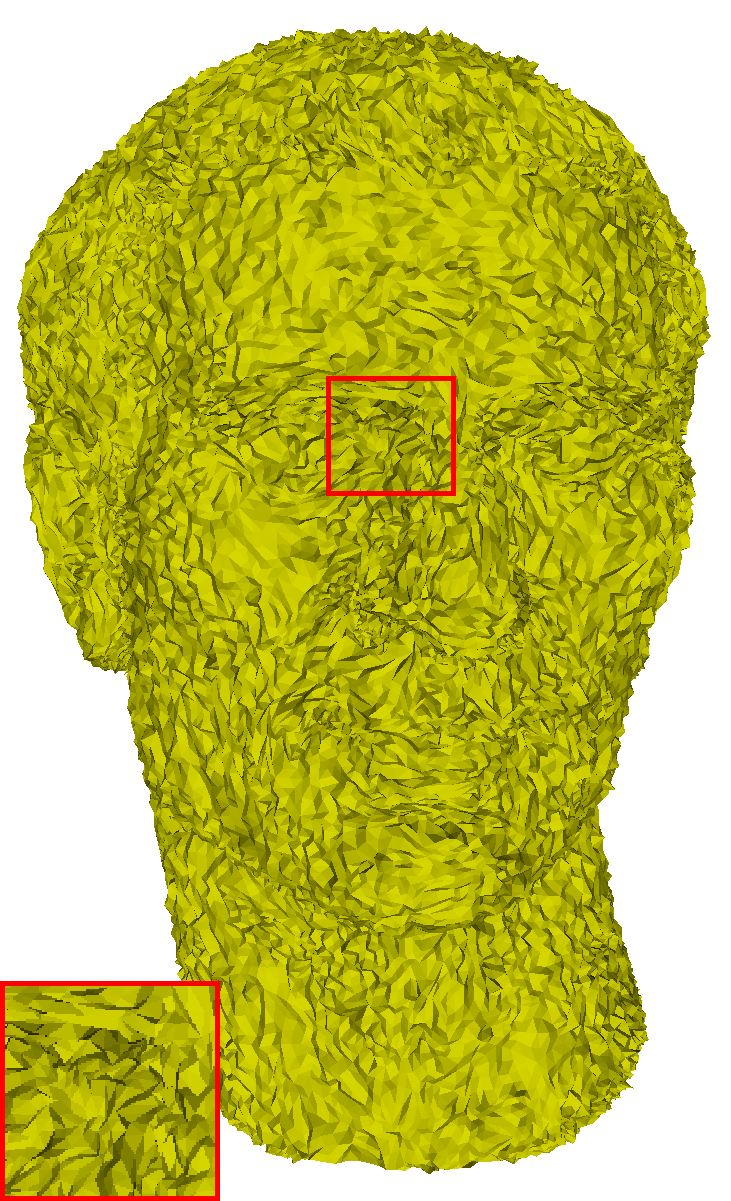}}
		{\includegraphics[width=0.2\textwidth]{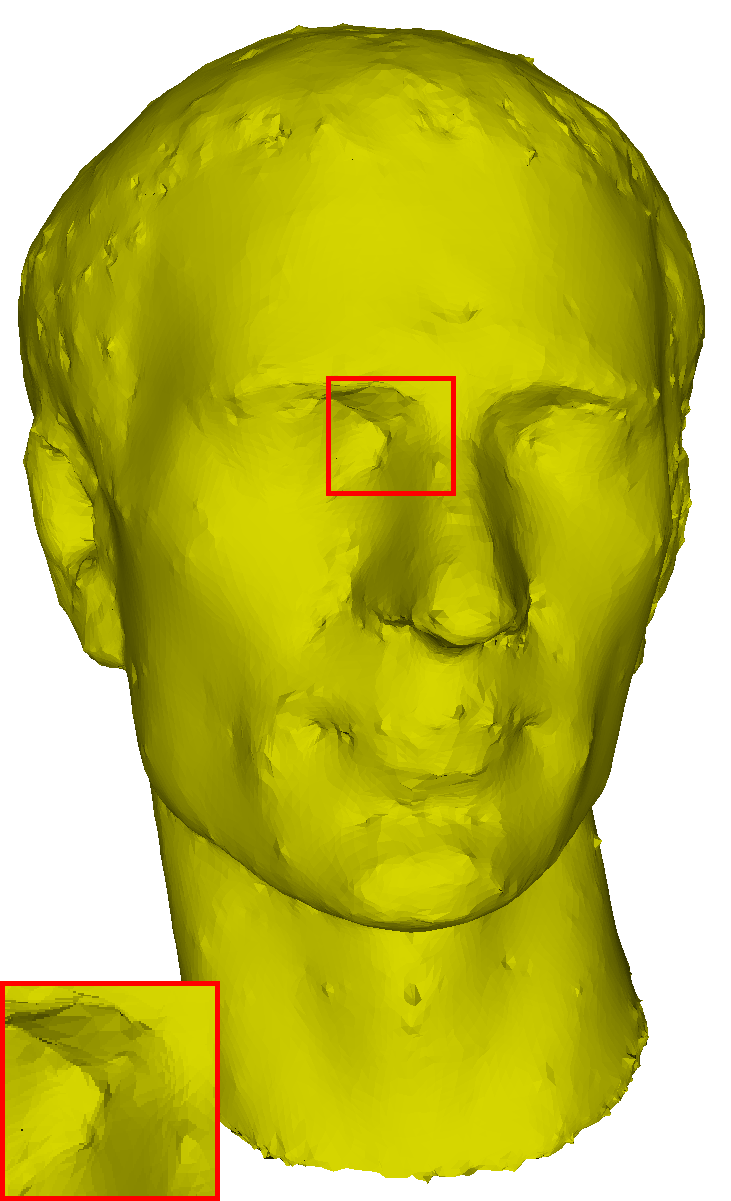}}
		{\includegraphics[width=0.2\textwidth]{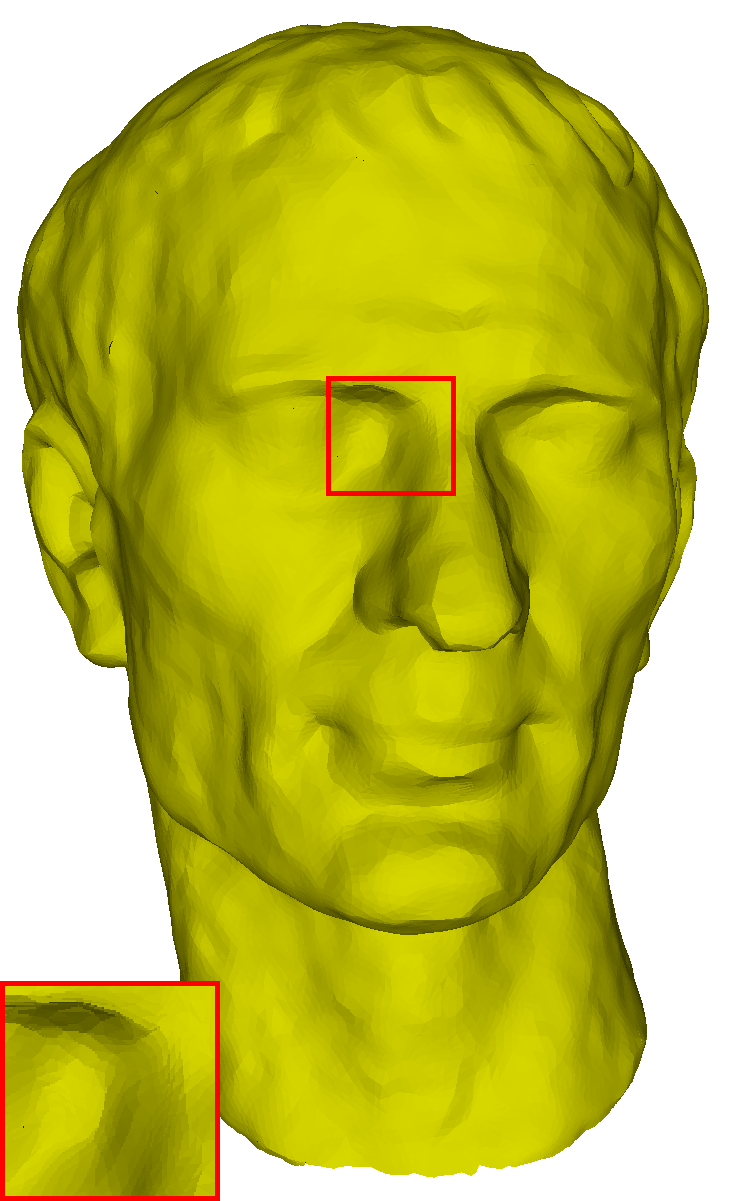}}
		{\includegraphics[width=0.2\textwidth]{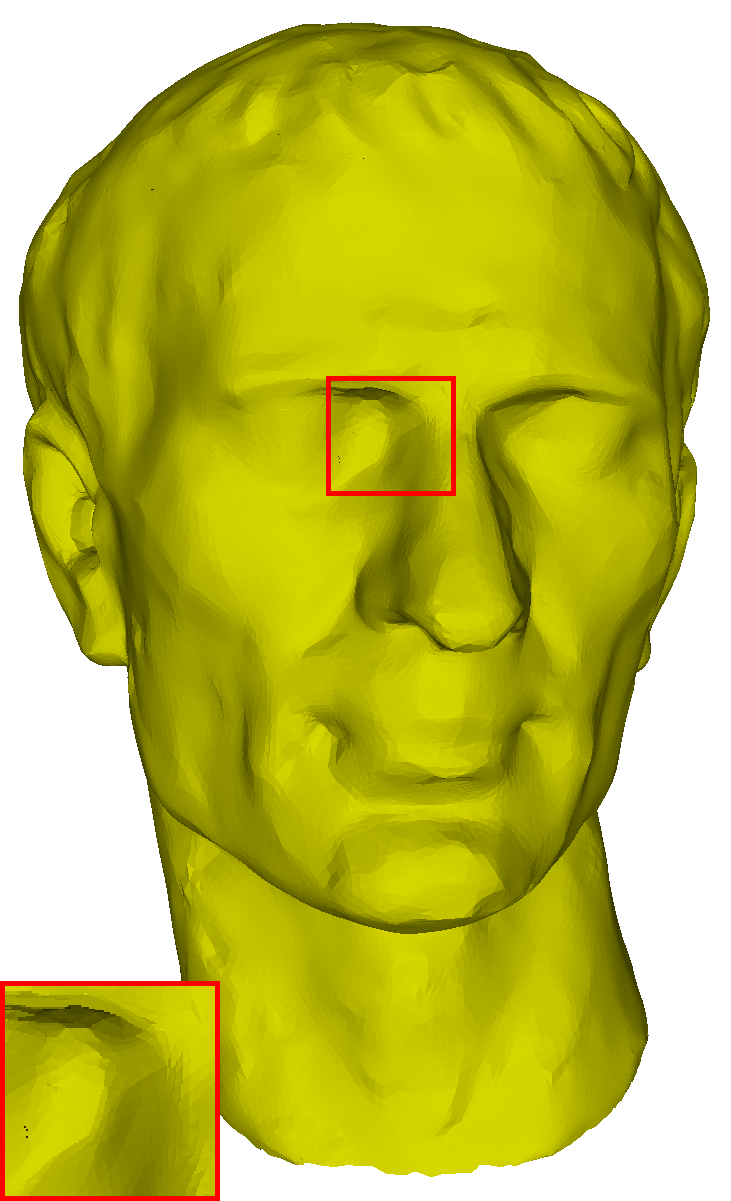}}
		
		{\includegraphics[width=0.2\textwidth]{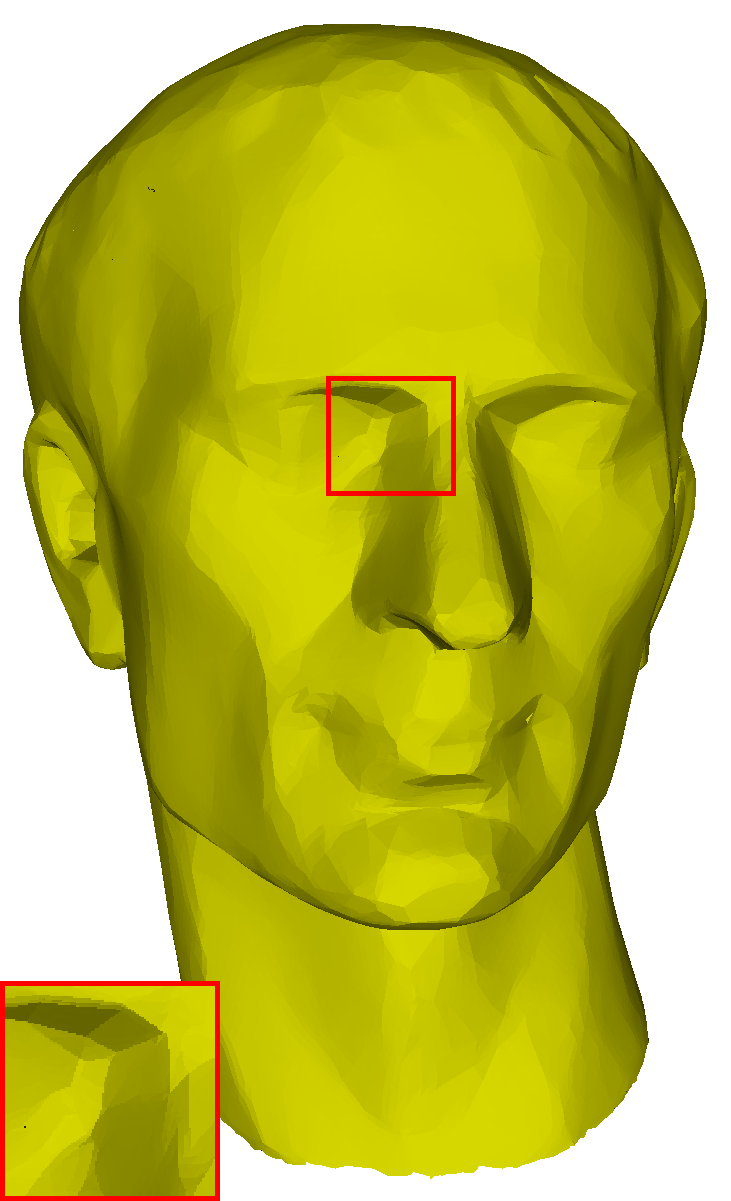}}
		{\includegraphics[width=0.2\textwidth]{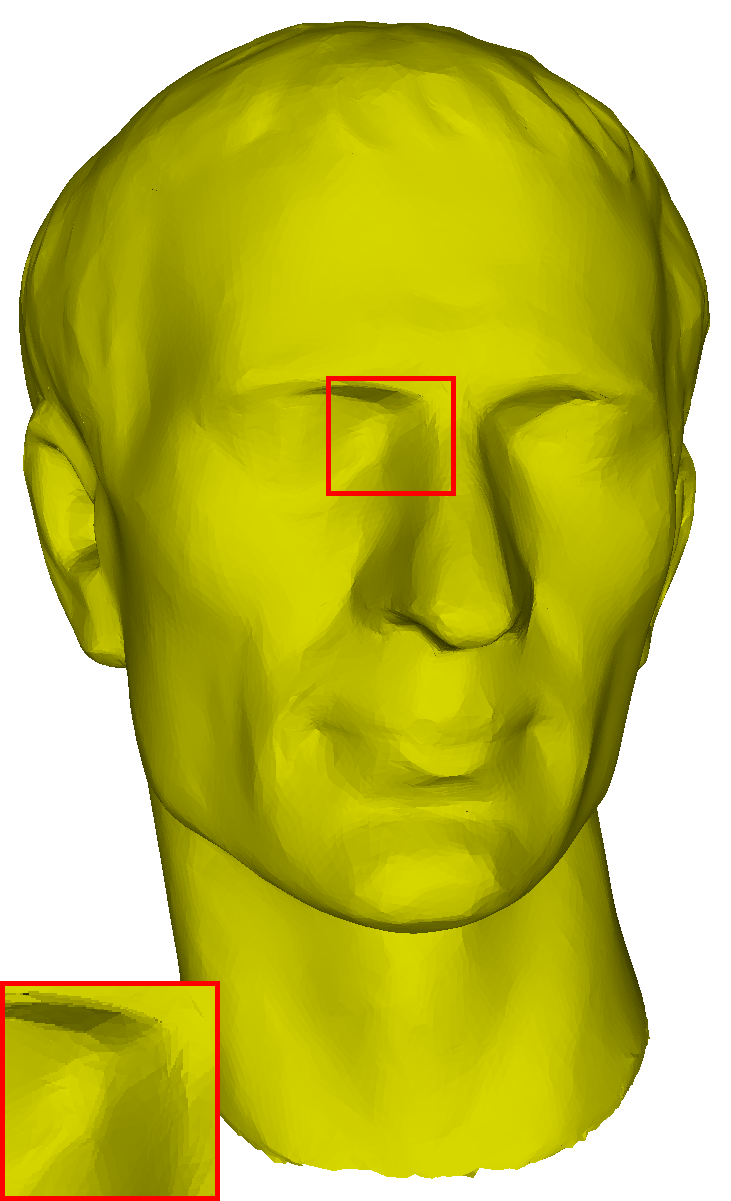}}
		{\includegraphics[width=0.2\textwidth]{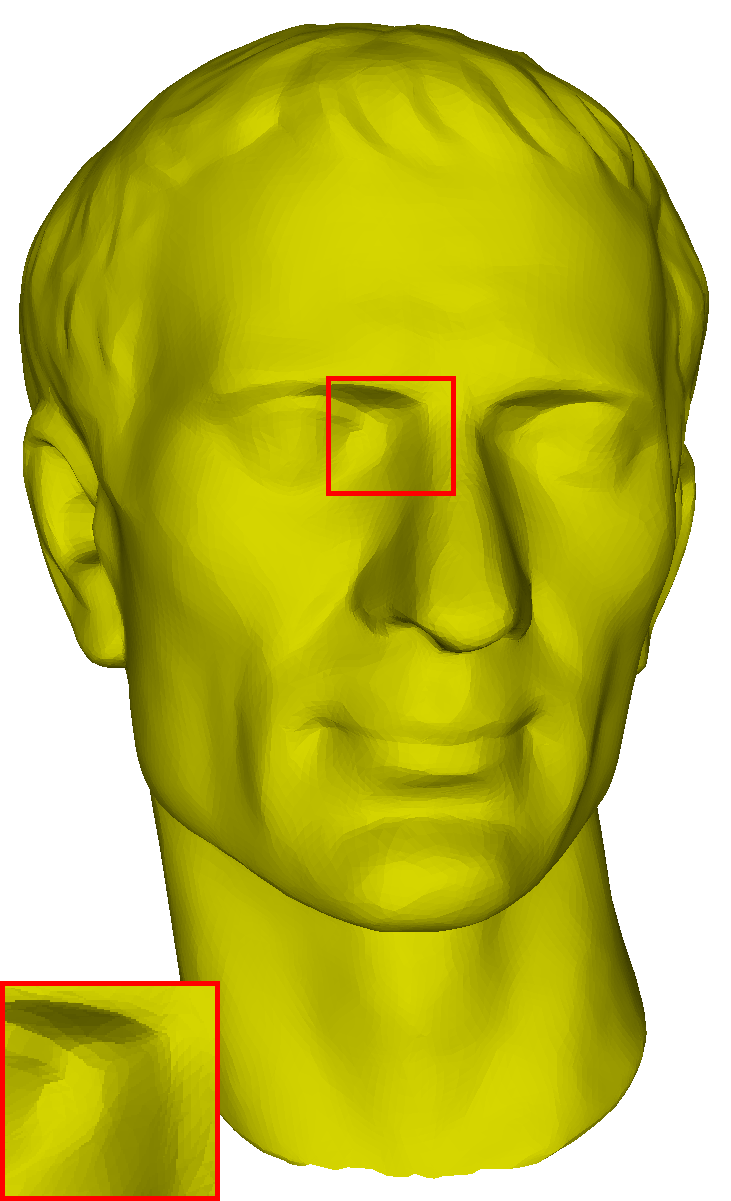}}
		{\includegraphics[width=0.2\textwidth]{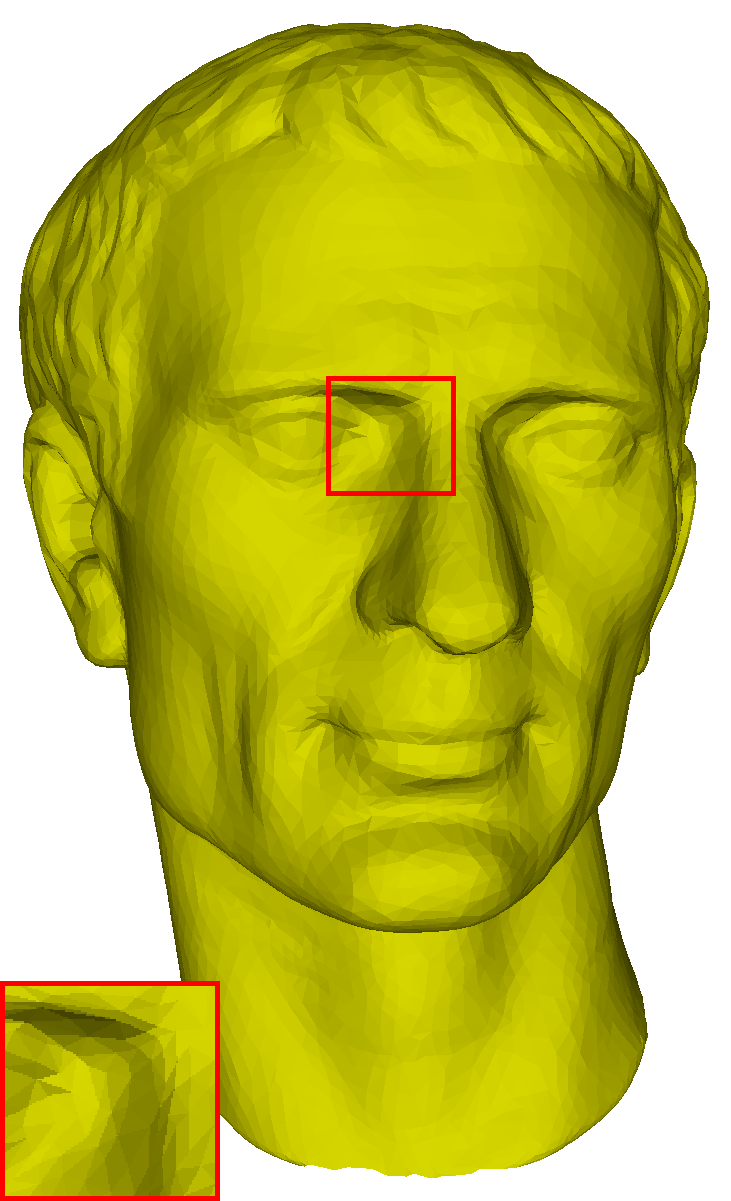}}
		
		\subfigure[Noisy]
		{\includegraphics[width=0.2\textwidth]{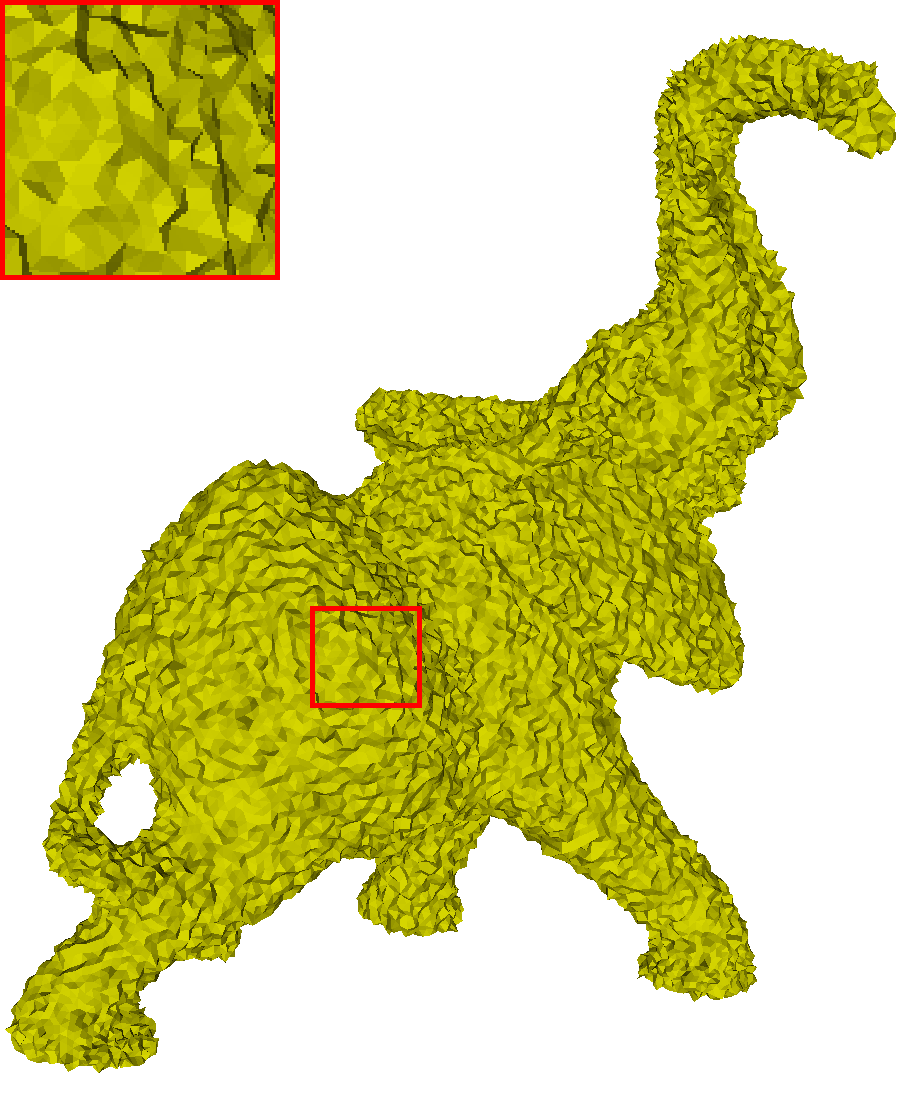}}
		\subfigure[BF\cite{fleishman2003bilateral}]
		{\includegraphics[width=0.2\textwidth]{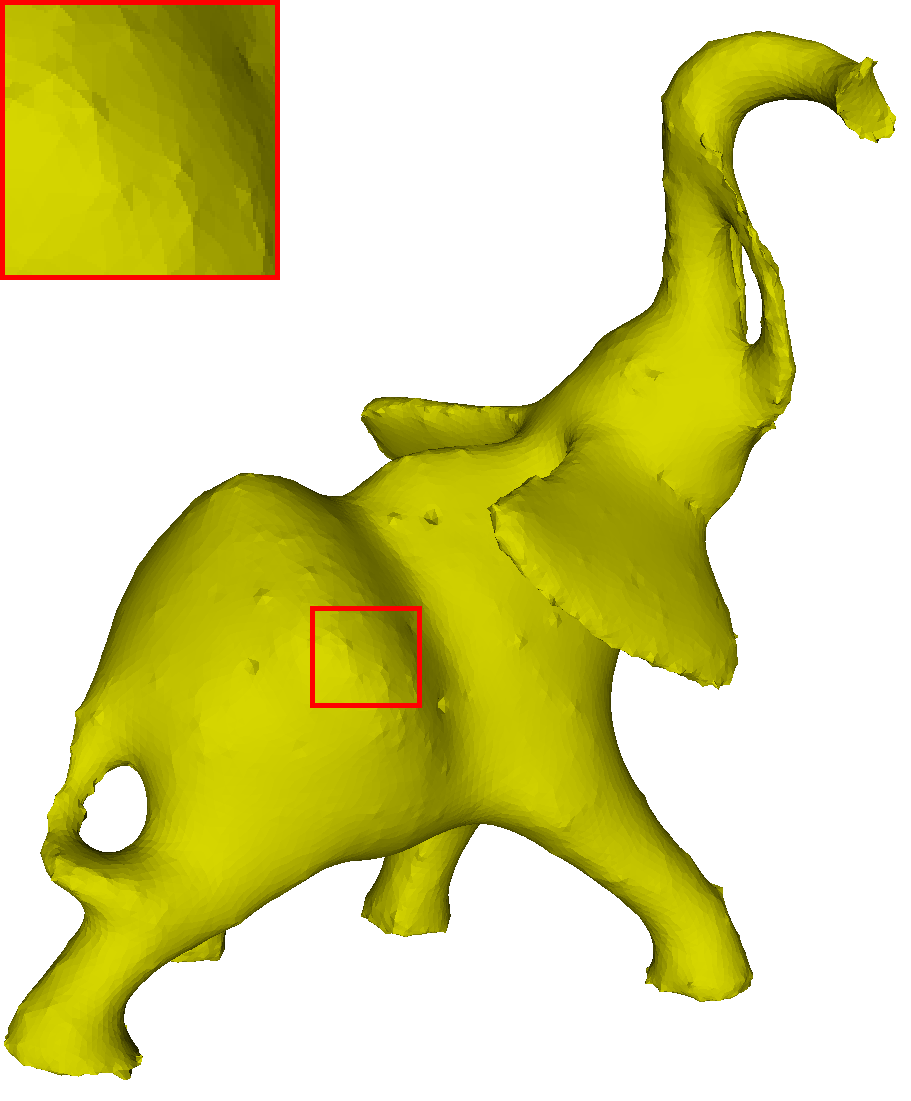}}
		\subfigure[GNF\cite{zhang2015guided}]
		{\includegraphics[width=0.2\textwidth]{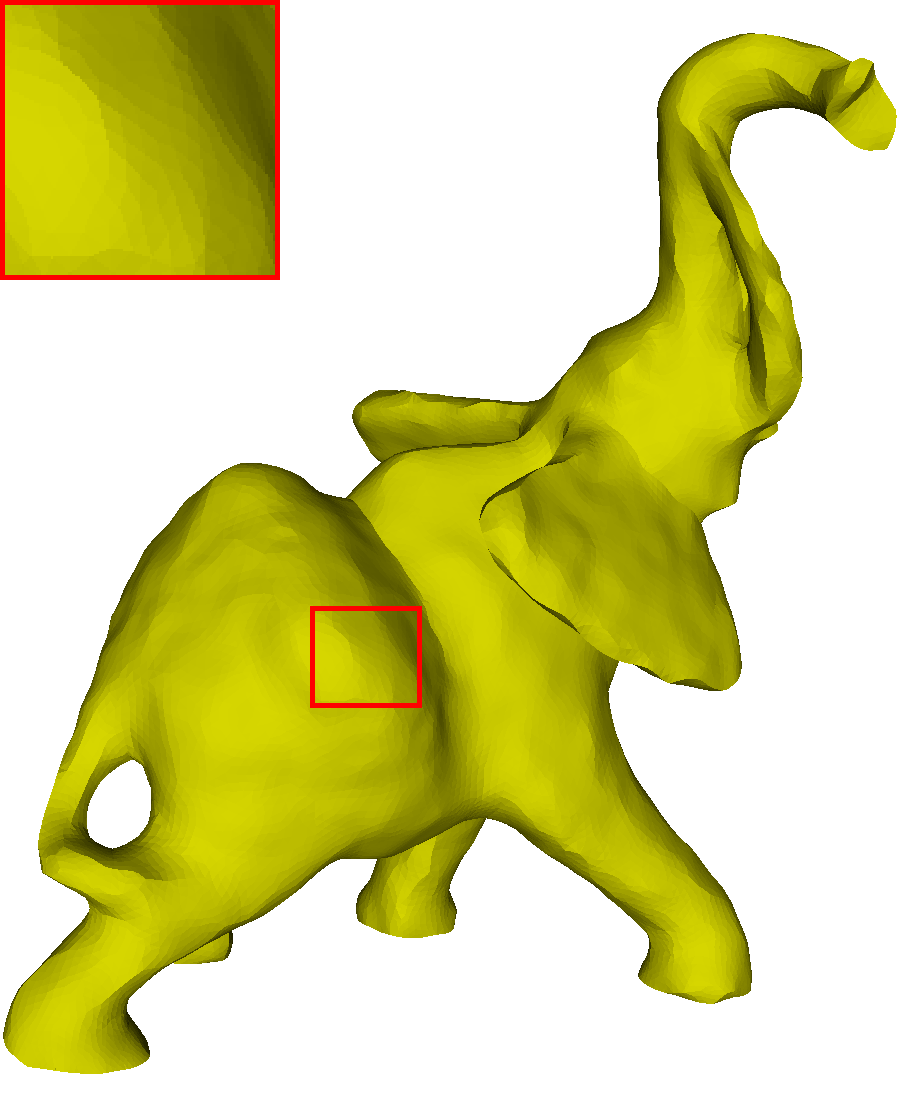}}
		\subfigure[CNR\cite{wang2016mesh}]
		{\includegraphics[width=0.2\textwidth]{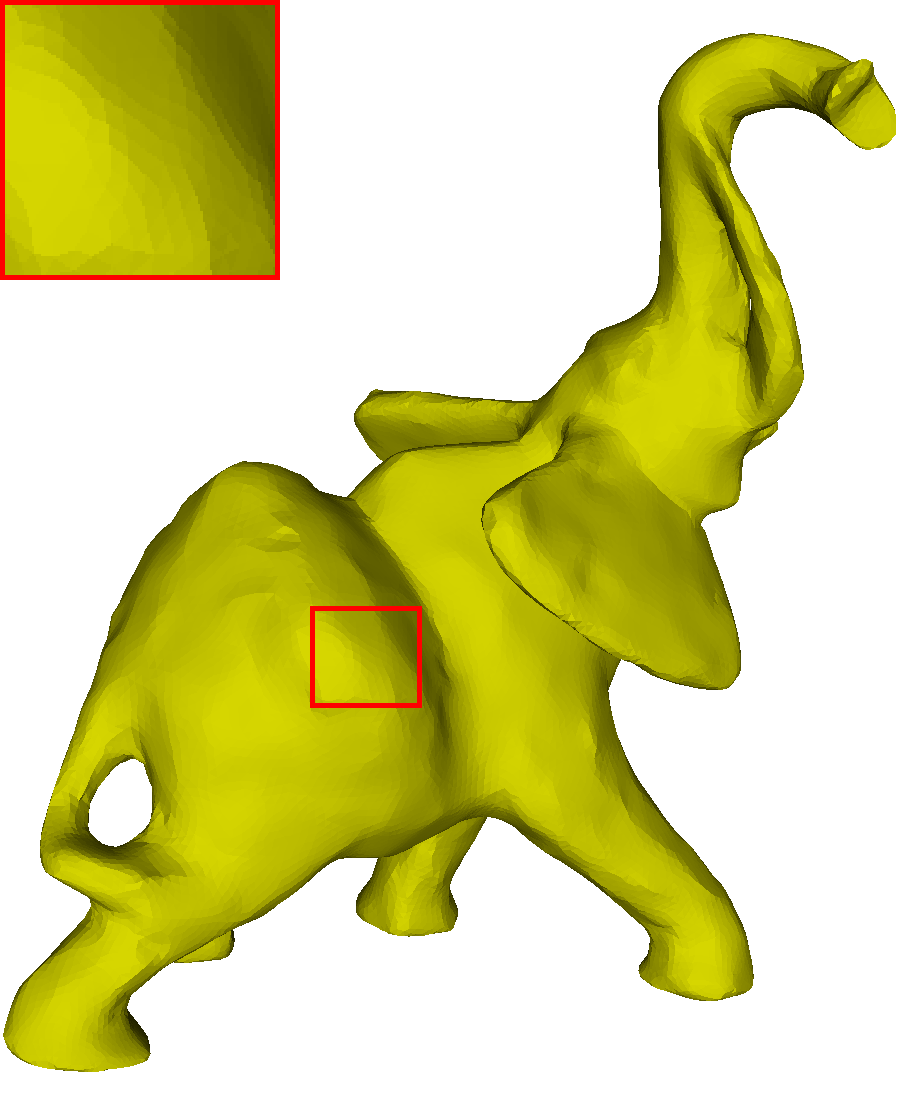}}
		
		\subfigure[$L_0$\cite{he2013mesh}]
		{\includegraphics[width=0.2\textwidth]{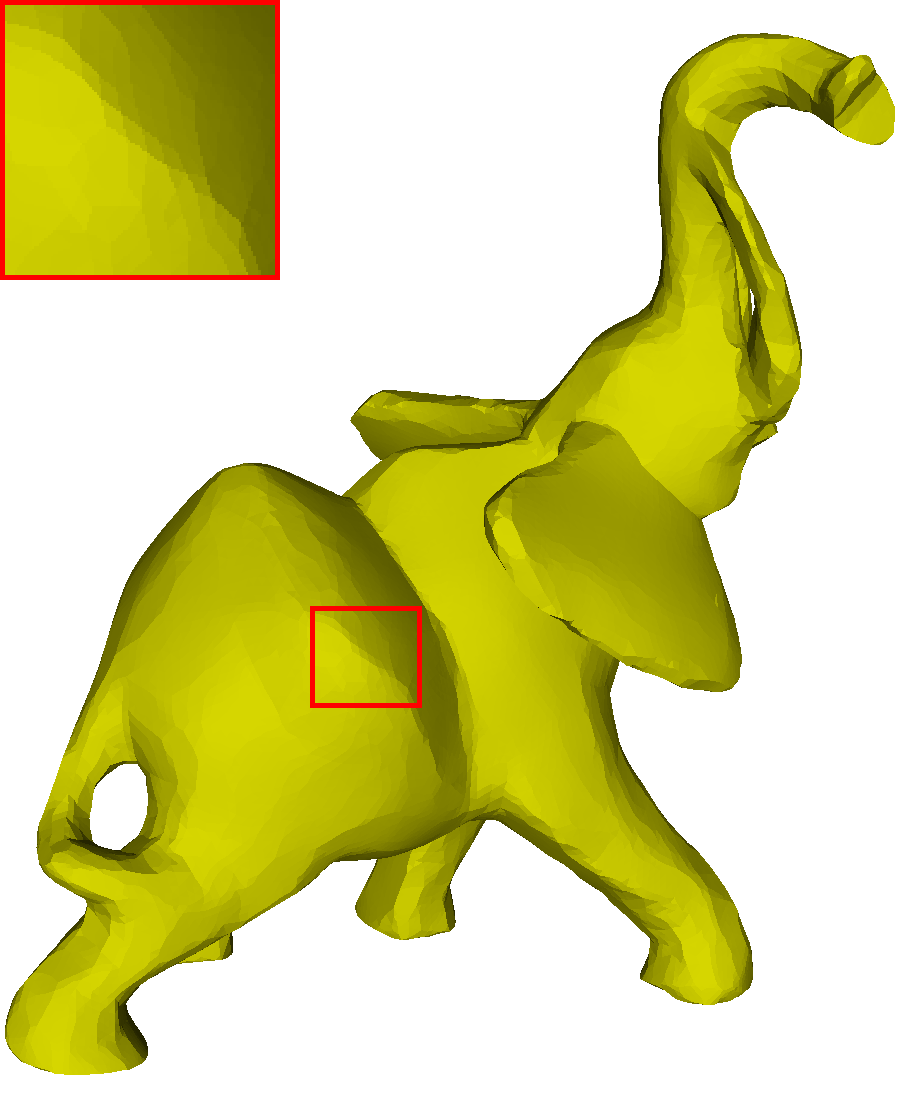}}
		\subfigure[TGV\cite{liu2021mesh}]
		{\includegraphics[width=0.2\textwidth]{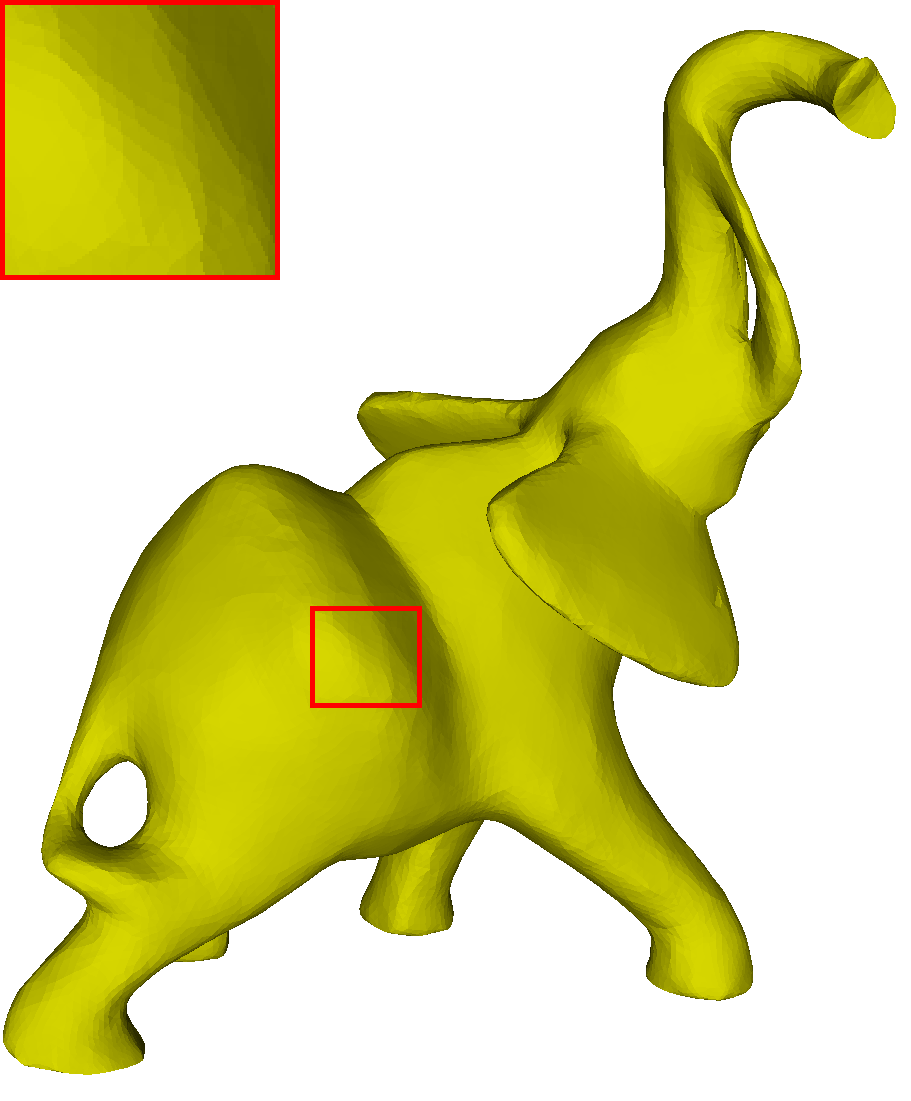}}
		\subfigure[Ours]
		{\includegraphics[width=0.2\textwidth]{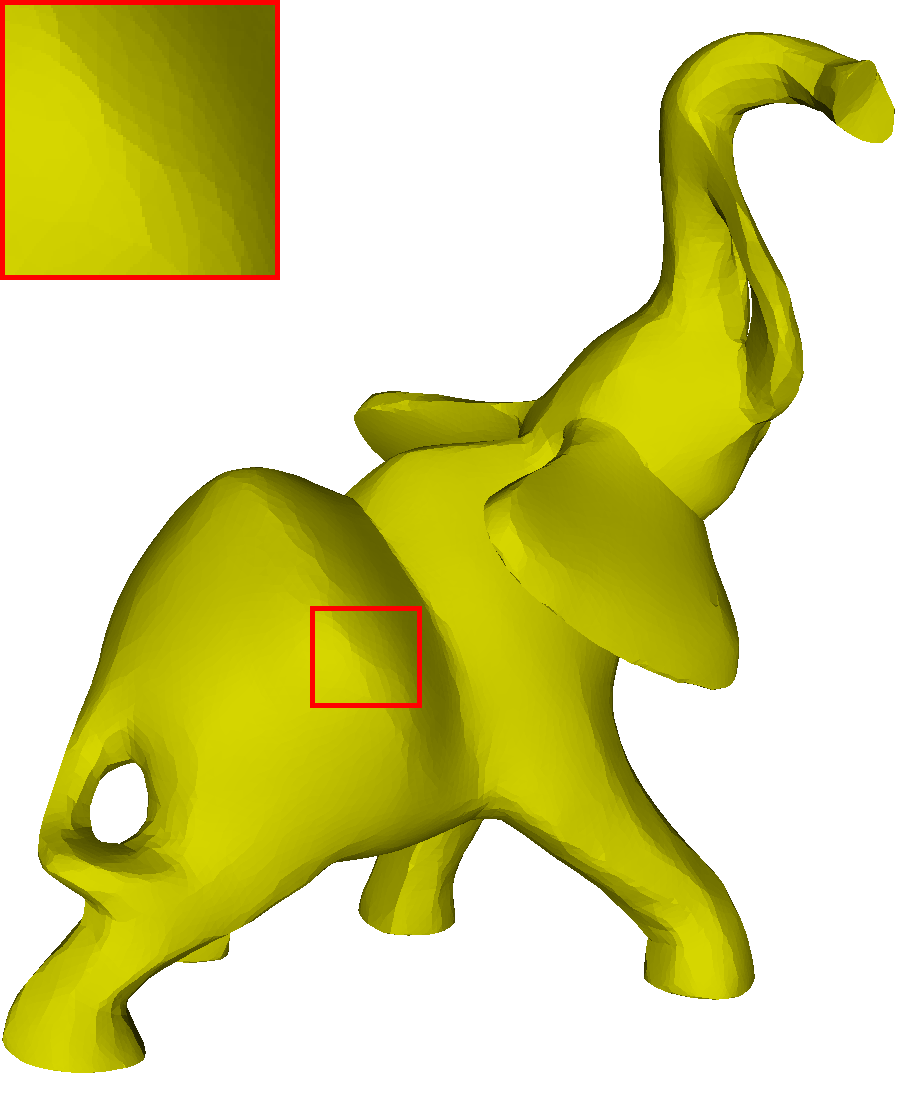}}
		\subfigure[GT]
		{\includegraphics[width=0.2\textwidth]{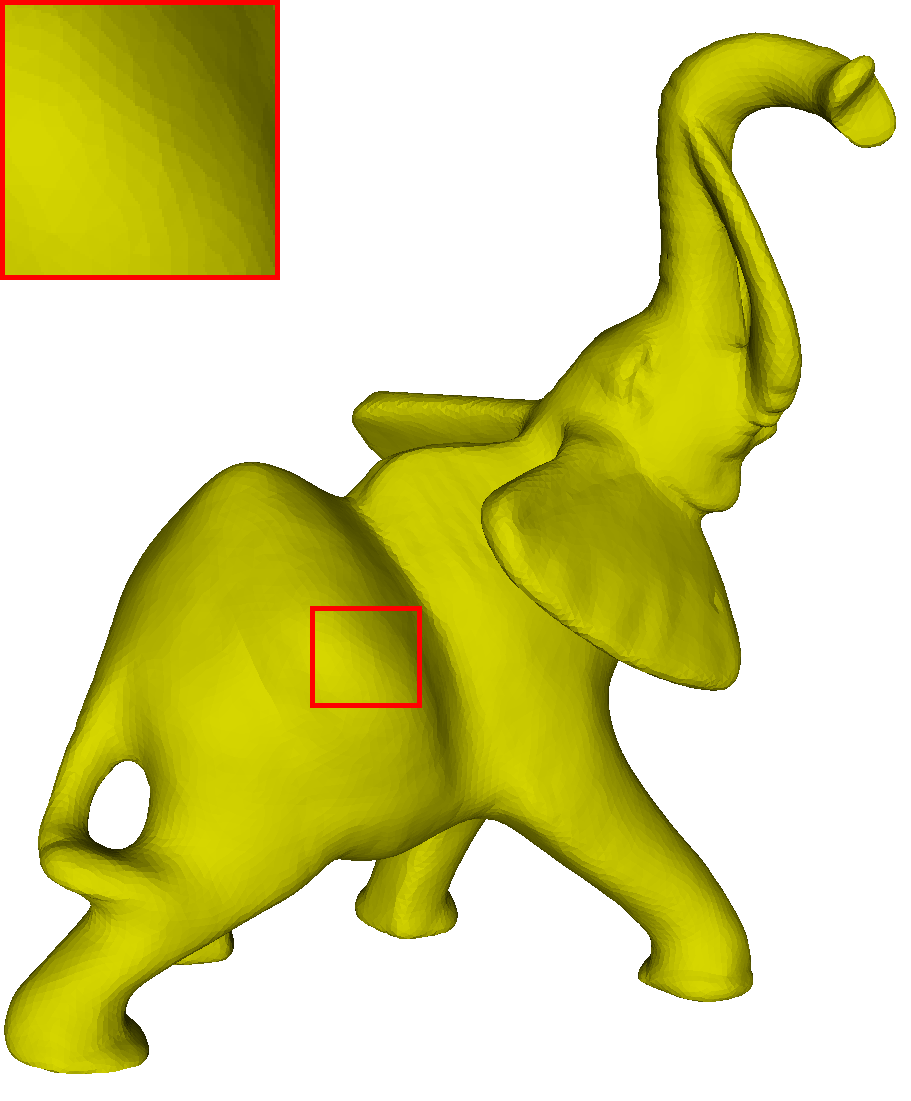}}
	\end{center}
	\caption{Mesh denoising results. (a) Input (Gaussian noise $\sigma = 0.3\bar{L}$), (b) Bilateral filter (BF)\cite{fleishman2003bilateral}, (c) Guided normal filter (GNF)\cite{zhang2015guided}, (d) Cascaded normal regression (CNR)\cite{wang2016mesh}, (e) $L_0$ minimization\cite{he2013mesh}, (f) TGV  regularization\cite{liu2021mesh}, (g) Our result and (h) Ground Truth (GT).} 
	\label{Fig:fig3}
\end{figure*}

\begin{figure*}[!t]
	\begin{center}
		{\includegraphics[width=0.2\textwidth]{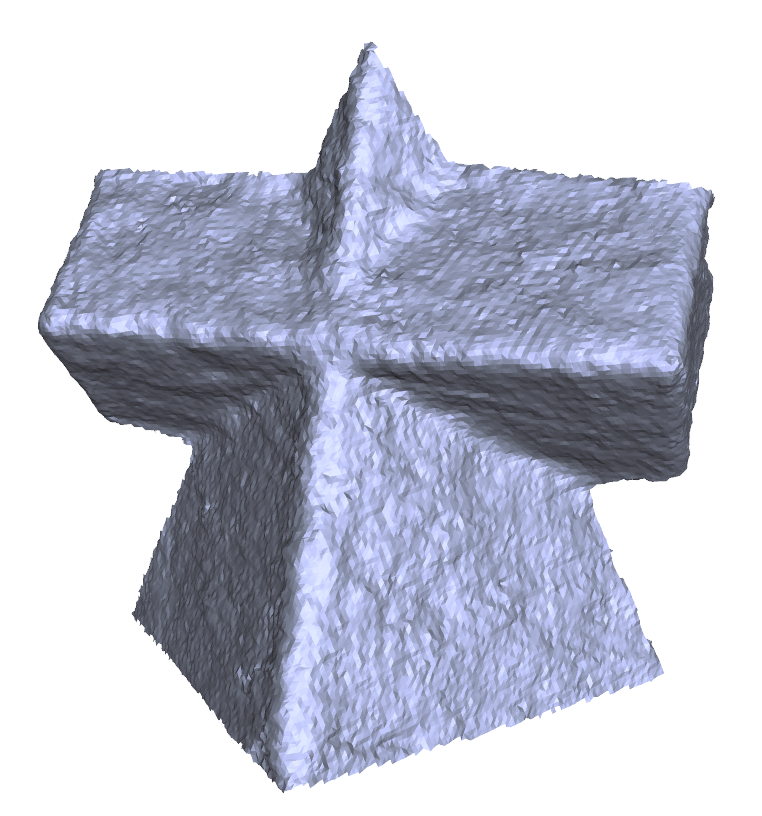}}
		{\includegraphics[width=0.2\textwidth]{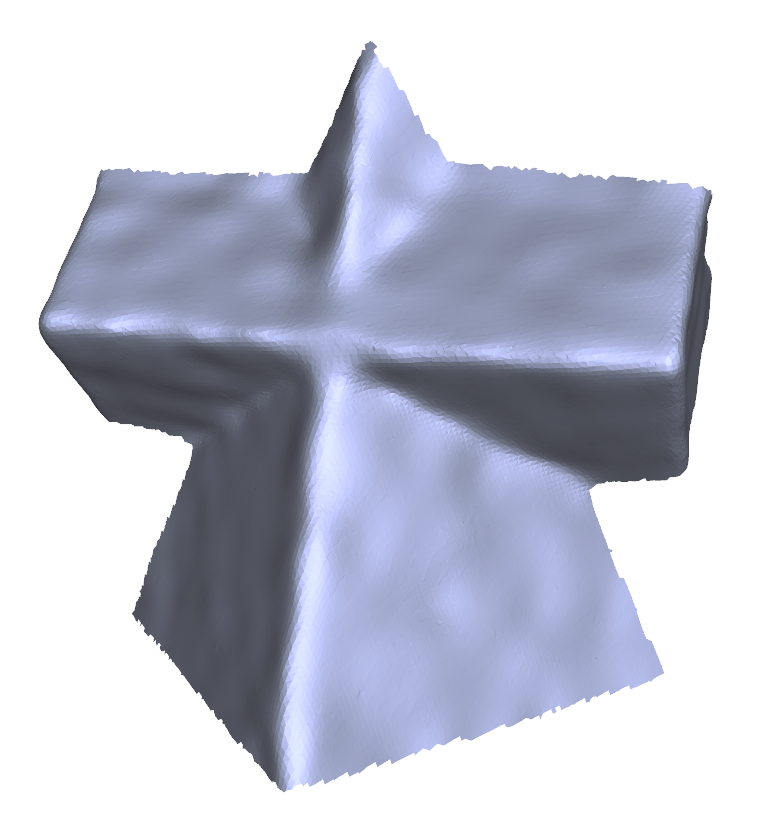}}
		{\includegraphics[width=0.2\textwidth]{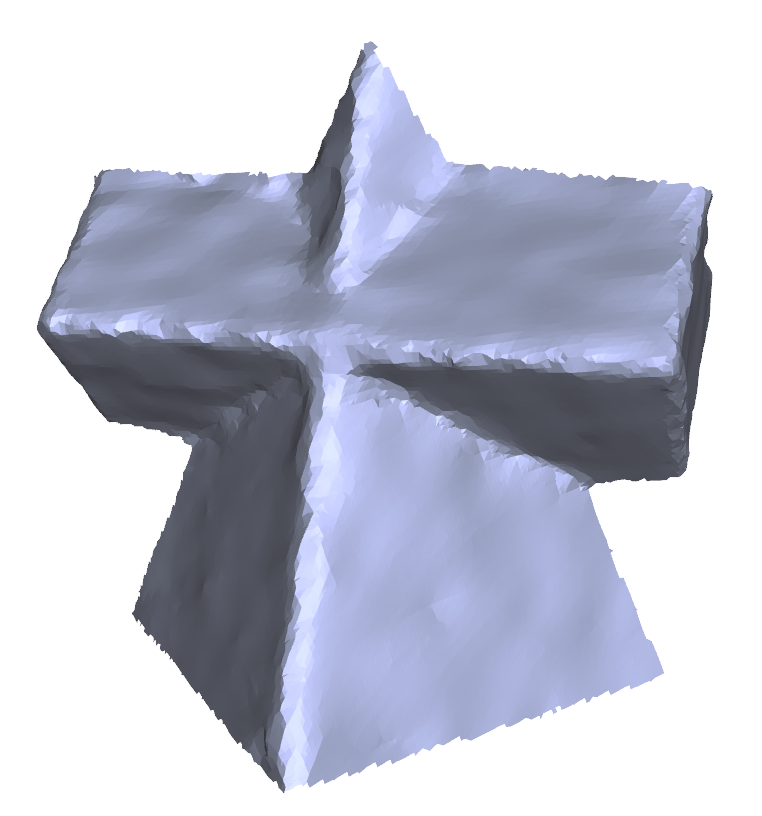}}
		{\includegraphics[width=0.2\textwidth]{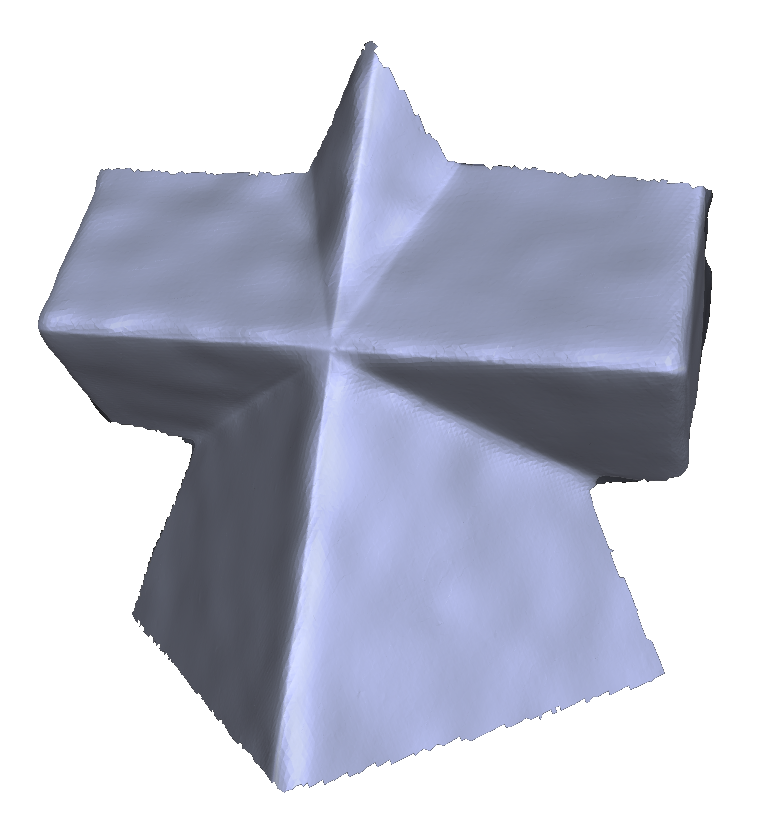}}
		\subfigure[]
		{\includegraphics[width=0.2\textwidth]{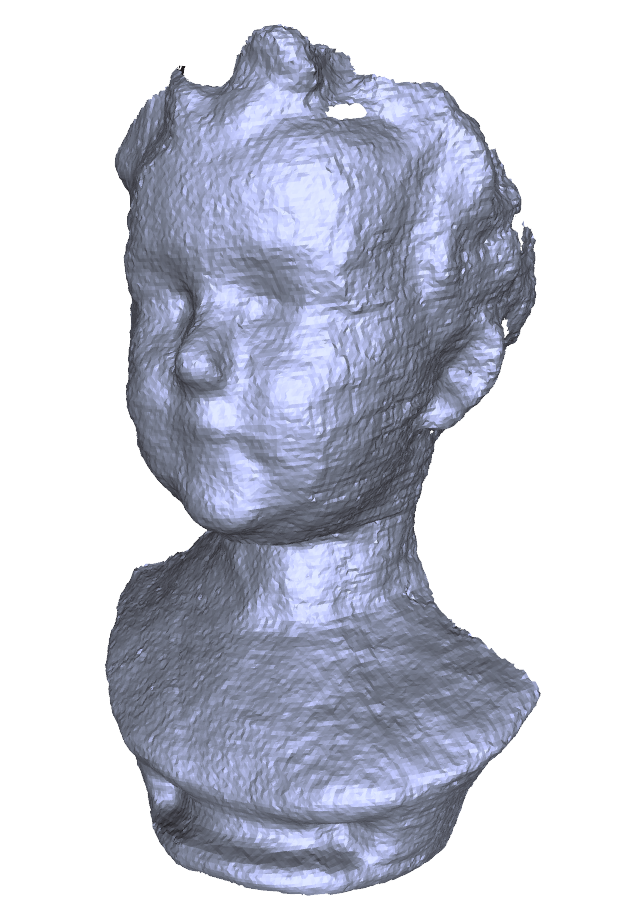}}
		\subfigure[]
		{\includegraphics[width=0.2\textwidth]{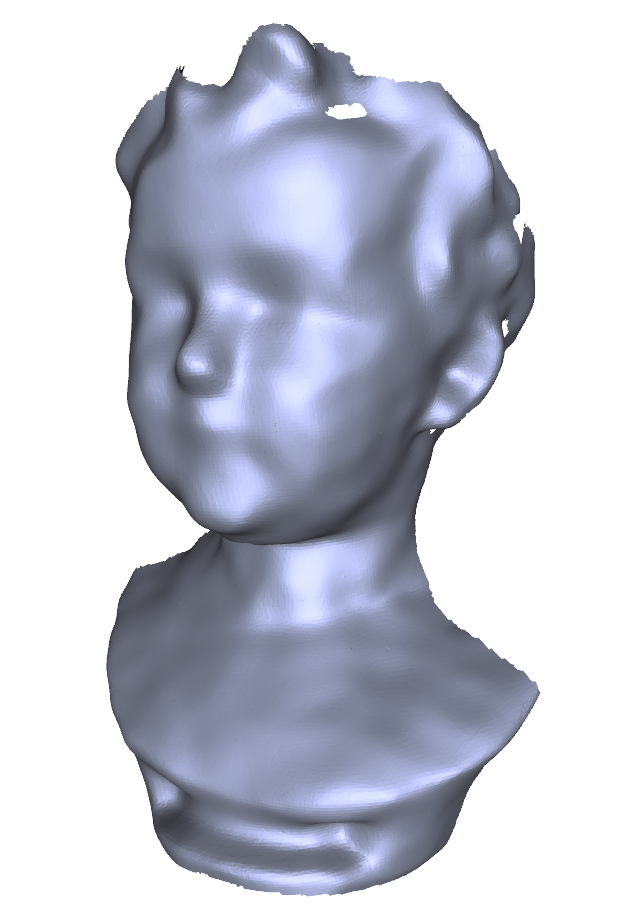}}
		\subfigure[]
		{\includegraphics[width=0.2\textwidth]{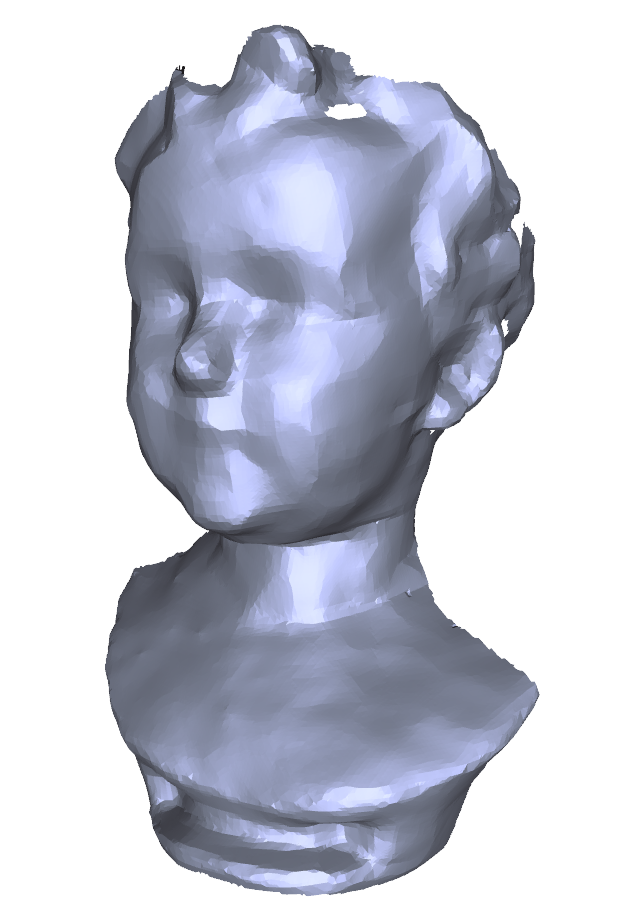}}
		\subfigure[]
		{\includegraphics[width=0.2\textwidth]{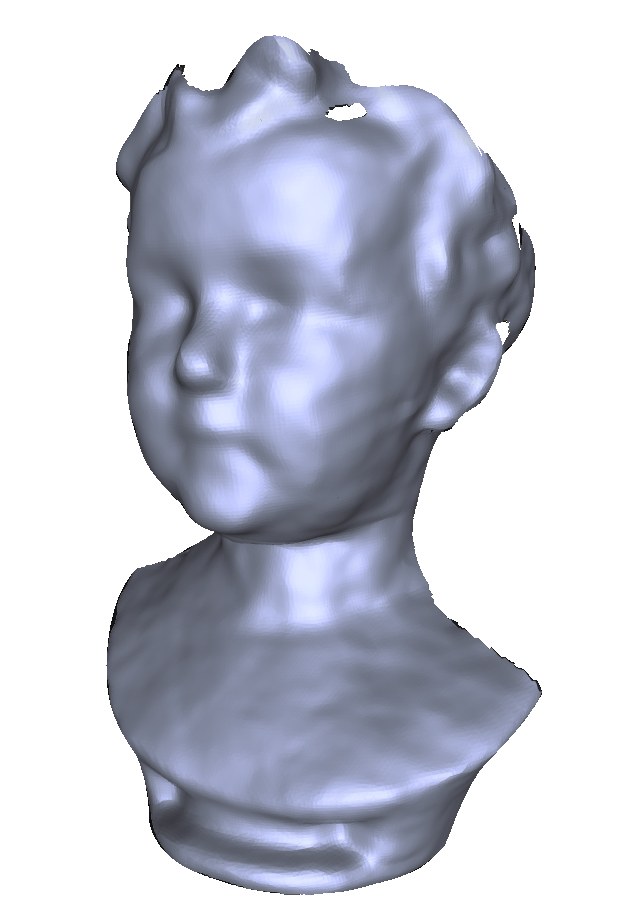}}
	\end{center}
	\caption{Mesh denoising on scanned surfaces. (a) Noisy input, (b) Cascaded normal regression~\cite{wang2016mesh}, (c) $L_0$ minimization~\cite{he2013mesh}, (d) Our result.}
	\label{Fig:fig4}
\end{figure*}

As shown in Fig. \ref{Fig:fig2}, the original surface contains corners, edges and polynomial smoothing surfaces. The BF method removes noise in smoothing areas but also seriously blurs sharp features; the GNF and CNR methods produce much better smoothing results but slightly blur the strong edges; and $L_0$ minimization retains sharp features but leads to slanted artifacts in the smoothing regions. While, our method not only produces a similar result as the cutting-edge high-order (HO) regularization\cite{liu2019triangulated} in polynomial-smoothing regions but also preserves the sharpening corners and edges. This is further demonstrated by the results in Fig. \ref{Fig:fig3} and the real scanned surfaces in Fig. \ref{Fig:fig4}. The experiments demonstrate the extension of our semi-sparse model to triangular meshes. We define the differential operator over the edge and only update the vertexes for mesh denoising. It is also possible to take a two-stage strategy for both vertexes and normal vectors as explained in \cite{liu2019triangulated}. We refer the interested reader to\cite{liu2019triangulated, liu2021mesh,  meyer2003discrete} for more details. 

\section*{Acknowledgements}
The research is financially supported by the Research Foundation – Flanders (FWO) Odysseus 1 under Grant G.0H94.18N; Methusalem Programme of the Ghent University Special Research Fund (BOF) under Grant 01M01021.

\end{document}